\documentstyle[aps,aps,epsf]{revtex}

\begin{document}
\newcommand\1{$\spadesuit$}
\newcommand\2{$\clubsuit$}
\draft
\twocolumn[\hsize\textwidth\columnwidth\hsize\csname 
@twocolumnfalse\endcsname
\title{The Trans-Planckian Problem of Inflationary Cosmology}

\author{J\'er\^ome Martin}
\address{DARC, Observatoire de Paris--Meudon, \\ 
UMR 8629 -- CNRS, 92195 Meudon Cedex, France. \\
e-mail: martin@edelweiss.obspm.fr
}
\author{Robert~H.~Brandenberger}
\address{Department of Physics, Brown University, Providence, RI 02912, USA.\\
e-mail: rhb@het.brown.edu
}

\date{April 3, 2000}
\maketitle

\begin{abstract}
In most current models of inflation based on a weakly self-coupled scalar matter 
field minimally coupled to gravity, the period of inflation lasts so long that, at 
the beginning of the inflationary period, the physical wavelengths of comoving 
scales which correspond to the present large-scale structure of the Universe were 
smaller than the Planck length. Thus, the usual computations of the spectrum of 
fluctuations in these models involve extrapolating low energy physics (both in the 
matter and gravitational sector) into regions where this physics is not applicable. In 
this article we study the dependence of the usual predictions of inflation for 
the spectrum of cosmological fluctuations on the hidden assumptions about super-Planck 
scale physics. We introduce a class of modified dispersion relations to mimic possible 
effects of super-Planck scale physics, and find that, given an initial state 
determined by minimizing the energy density, for dispersions relations introduced 
by Unruh the spectrum is unchanged, whereas for a class of dispersion 
relations similar to those used by Corley and Jacobson (which 
involve a more radical departure from the usual linear relation) important deviations 
from the usual predictions of inflation can be obtained. Some implications of 
this result for the unification of fundamental physics and early Universe 
cosmology are discussed.
\end{abstract}

\vspace*{1cm}
]

\section{Introduction}

The inflationary Universe scenario \cite{Guth} is the first theory of the very early 
Universe to provide a mechanism \cite{MCL} for the production of density 
fluctuations on scales of cosmological interest based on causal physics (see 
also Ref. \cite{Press} for initial ideas). The key point is that during the period 
of inflation fixed comoving scales are stretched exponentially compared to the Hubble 
radius. Thus, the wavelengths corresponding to the present large-scale structure in 
the Universe and to the measured Cosmic Microwave Background (CMB) anisotropies were 
equal to the Hubble radius about 50 Hubble expansion times before the end of 
inflation. This gives rise to the possibility that causal physics acting before that 
time can generate fluctuations on these scales while they are of sub-Hubble length.
\par
Most current models of inflation are based on weakly self-coupled scalar matter 
fields minimally coupled to gravity. In this context, quantum vacuum fluctuations 
provide \cite{MCL} a causal mechanism for generating fluctuations. In fact, the 
coupled linear metric and matter fluctuations can be quantized in a unified 
manner \cite{SM}. The problem reduces to the quantization of a free scalar field with 
a time-dependent mass (see e.g. Ref. \cite{MFB} for a comprehensive review). An 
initial vacuum state thus undergoes squeezing during inflation, and this leads to the 
generation of fluctuations. According to the standard calculations 
\cite{MCL,S82,H82,GP82,BST,RB84}, the predicted spectrum is scale-invariant (modulo 
a mild deviation from scale-invariance which stems from the time-dependence 
of the Hubble constant during the inflationary period).
\par
There are good heuristic reasons \cite{Press} to expect a scale-invariant spectrum of 
fluctuations to emerge from inflation. Since de-Sitter space is 
time-translation-invariant, one should expect the amplitude of the density 
fluctuations ${\delta M}/M$ to be independent of the scale (labelled by the comoving 
wavenumber $n$) if measured at the time when the corresponding wavelength crosses the 
Hubble radius $l_{\rm H}$ during the inflationary period. Since microphysics cannot 
change the physical amplitude of the mass fluctuations while the wavelength is larger 
than $l_{\rm H}$, one therefore expects ${\delta M}/M$ to be independent of $n$ when 
measured at the time $t_{\rm f}(n)$ when the scale re-enters the Hubble radius in the 
post-inflationary Friedmann-Robertson-Walker period:
\begin{equation}
{{\delta M} \over M}\bigl(n, t_{\rm f}(n)\bigr) \, = \, {\rm const.} \, ,
\end{equation}
which is the definition of a scale-invariant Harrison-Zel'dovich 
spectrum \cite{HZ}.
\par 
The time-translation-invariance is, however, broken in the current models of 
inflation. The calculations are done by picking an initial time $t_{\rm i}$ (e.g. the 
beginning of the inflationary period), by choosing a specific state of the quantum 
fields at this time (e.g. the {\it local Minkowski vacuum state} \cite{RB84} or the 
Bunch-Davies vacuum \cite{BD}), by evolving this state using the linearized equations 
of motion, and by finally calculating the correlation functions and expectation 
values of interest. In this context, the emergence of a scale-invariant spectrum of 
fluctuations is seen to arise from a subtle cancellation of the 
wavenumber dependence in the 
initial state wave function and in the growth factor before Hubble radius crossing, 
and thus depends explicitly on the initial state chosen. States can be 
found \cite{MRS99} which do not yield a scale-invariant spectrum. Thus, it is clear 
that the prediction of a Harrison-Zel'dovich spectrum is not completely generic in 
current models of inflation. 
\par
There is, however, a much more serious potential problem for the claim that current 
models of inflation based on weakly self-coupled scalar fields generically lead to 
a scale-invariant spectrum of fluctuations. Most of these models of inflation 
involve (see e.g. \cite{Linderev} for a recent review) a period of inflation much 
longer than the 60 e-foldings of inflation required to solve the horizon and flatness 
problems of standard cosmology. Since wavelengths exponentially redshift during 
inflation, the physical wavelengths of the modes which correspond to the present 
large-scale structure in the Universe were, in those models, much smaller than the 
Planck length at the initial time $t_{\rm i}$. Thus, the usual computations of the spectrum 
of fluctuations involve extrapolating weakly self-coupled field theory coupled to 
classical gravity into a regime where these theories are known to break down. 
\par
This problem is analogous to the Trans-Planckian problem for black hole
physics (see Ref. \cite{TJ00} for a recent overview). In black hole physics there is 
an arbitrarily large blue shift when following modes of Hawking radiation at future 
infinity into the past, and the usual calculations of Hawking 
radiation \cite{Hawking} seem suspect (see e.g. Ref. \cite{TJ91} for a discussion 
of this point).
\par
In the case of the black hole problem, it was recently shown by 
Unruh \cite{Unruh}, Brout et al. \cite{Brout}, Hambli and Burgess \cite{HB} 
and by Corley and Jacobson \cite{CJ} that 
the prediction of a thermal Hawking spectrum of black hole radiation is insensitive to 
modifications of the physics at the ultraviolet end of the spectrum. In these works, the 
dispersion relation of the quantum fields was modified (in rather ad-hoc ways) at 
energies larger than some ultraviolet scale $k_{\rm C}$, and it was found that the 
spectrum of radiation at future infinity at wavenumbers much smaller than $k_{\rm C}$ is 
insensitive to the modifications considered. In this sense, Hawking radiation from 
black holes was shown to be an infrared effect.
\par
The obvious question is whether a similar conclusion will hold for the generation of 
fluctuations in inflationary cosmology. This is the question we will address in this 
paper. We will consider a free scalar field in an inflationary background [de Sitter 
phase of a Friedmann-Robertson-Walker cosmology with scale factor $a(t)$]. This scalar 
field can represent the scalar metric fluctuations, the gravitational wave mode, or a 
matter scalar field on the fixed background geometry - the case of most interest for 
cosmology corresponds to scalar metric fluctuations. We will modify the usual 
dispersion relation 
\begin{equation}
\omega ^2=k^2, \quad k^2 \equiv {{n^2} \over {a^2}} ,
\end{equation}
where $n$ and $k$ are the comoving and physical wavenumbers, respectively, for 
values of $k$ larger than some cutoff scale $k_{\rm C}$, and will calculate the 
predicted spectrum of fluctuations in the modified theory for well-motivated initial 
quantum states, states which in the unmodified theory coincide with the state usually 
chosen as the initial state. The modified dispersion relations which we use are the 
same as the ones used by Unruh \cite{Unruh} and by Corley and Jacobson \cite{CJ}. As 
preferred initial states we will use either the state which minimizes the energy 
density at the initial time $t_{\rm i}$, following the approach of Brown and 
Dutton \cite{Brown78}, or a naive generalization of the local Minkowski vacuum.
\par
We find that in the case of Unruh's dispersion relation, the spectrum of density 
fluctuations is unchanged in the {\it minimum energy density} initial state. However, 
in the case of the family of dispersion relations generalizing the choice of Corley 
and Jacobson, the choice of the {\it minimum energy density} initial state leads to 
a spectrum of fluctuations which, depending on the specific member 
of the family of dispersion relations chosen, may be characterized by a 
tilt, by an exponential factor, and by superimposed oscillations. 
\par
Our work indicates that the prediction of a scale-invariant spectrum in inflationary 
cosmology depends sensitively on hidden assumptions about super-Planck-scale 
physics. This has important implications for the attempts to unify fundamental physics 
and early Universe cosmology. It is now a rather nontrivial question under which 
conditions a unified theory of all forces such as string or M-theory will lead to a 
scale-invariant spectrum, assuming for the moment that it does indeed lead to a 
period of inflation. 
\par
The outline of this paper is as follows. In Section II we demonstrate that the growth 
of linear density fluctuations, gravitational waves and linear scalar matter 
fluctuations can all be described in terms of the same framework: that of a free 
scalar field with a time-dependent mass. In Section III we introduce the two classes 
of modified dispersion relations which will be used in the calculations. The 
quantization of the scalar field in the time-dependent background and the construction 
of the {\it minimum energy density} initial state are reviewed in 
Section IV. Section V contains our calculations for both classes of 
dispersion relations. Our results are summarized and discussed in the final section.   

\section{Equivalence between cosmological perturbations and a 
fictitious scalar field}

Without loss of generality, the line element for the spatially flat 
Friedmann-Lema{\^\i}tre-Robertson-Walker (FLRW) background plus the 
perturbations can be written in the synchronous gauge according to \cite{LK,G}:
\begin{eqnarray}
\label{metricsg}
{\rm d}s^2 &=&
a^2(\eta )\biggl\{-{\rm d}\eta ^2+ \biggl[\delta _{ij}
+h(\eta ,{\bf n})Q\delta _{ij}\nonumber \\ 
& & +h_l(\eta ,{\bf n})\frac{Q_{,i,j}}{n^2}+
h_{\rm gw}(\eta ,{\bf n})Q_{ij}\biggr]{\rm d}x^i{\rm d}x^j\biggr\}.
\end{eqnarray}
In this equation, the dimensionless quantity ${\bf n}$ is the comoving 
wavevector related to the physical wavevector ${\bf k}$ through the 
relation ${\bf k}\equiv {\bf n}/a(\eta )$. $\eta $ is the conformal time 
related to the cosmic time $t$ by ${\rm d}t=a(\eta ){\rm d}\eta $. The functions 
$h$ and $h_l$ represent the 
scalar sector and $Q(x^i)$ is the eigenfunction of the Laplace 
operator on the flat spacelike hypersurfaces. The function $h_{\rm gw}$ 
represents the gravitational 
waves and $Q_{ij}(x^i)$ is the eigentensor of the Laplace operator. 
It is traceless and transverse, namely $Q_i{}^i=Q_{ij}{}^{,j}=0$. It 
is convenient to introduce the background quantity $\gamma (\eta )$ 
defined by $\gamma \equiv -\dot{H}/H^2$, where a dot means differentiation 
with respect to cosmic time and $H$ is the Hubble rate, 
$H\equiv \dot{a}/a$. We can also write $\gamma =1-{\cal H}'/{\cal H}^2$, where 
${\cal H}\equiv a'/a$ and a prime denotes differentiation with 
respect to the conformal time.
\par
In the tensor sector, we define the quantity $\mu _{\rm T}$ 
by $h_{\rm gw}\equiv \mu _{\rm T}/a$. Then, the equation of motion is 
given by \cite{Ggw}:
\begin{equation}
\label{eomtensor}
\mu _{{\rm T}}''+\biggl[n^2-\frac{a''}{a}\biggr]\mu _{{\rm T}}=0.
\end{equation} 
Since gravitational waves do not couple to matter, the last equation 
is valid for every type of matter. 
\par
In the scalar sector, it is convenient to work with a residual gauge invariant 
variable $\mu _{\rm S}$ defined by $\mu _{\rm S}\equiv [a/({\cal H} \sqrt{\gamma })]
(h'+{\cal H}\gamma h)$ where we have supposed $\gamma \neq 0$. The case 
$\gamma =0$ must be treated separately (see below). The quantity  
$\mu _{\rm S}$ is related 
to the gauge invariant Bardeen potential by $\Phi _{\rm B}^{({\rm SG})}
=[{\cal H}\gamma /(2n^2)][\mu _{\rm S}/(a\sqrt{\gamma })]'$ where the 
subscript `${\rm SG}$' means `calculated in the synchronous gauge' \cite{MS}. Therefore, 
knowing the solution for $\mu _{\rm S}$ permits the calculation of the Bardeen 
variable. If matter 
is described by a scalar field (the inflaton), then one can show 
that $\mu _{\rm S}$ obeys the equation:
\begin{equation}
\label{eomscalar}
\mu _{{\rm S}}''+\biggl[n^2-
\frac{(a\sqrt{\gamma   })''}{(a\sqrt{\gamma })}\biggr]\mu _{{\rm S}}=0.
\end{equation}
The case $\gamma =0$ corresponds to a scale factor 
$a(t)\propto e^{Ht}$, i.e. to the de Sitter manifold. Then, one can show 
that the exact solution to the 
perturbed Einstein equations is $\Phi _{\rm B}=0$: there are no density 
perturbations at all. This is because when the equation of state is 
$p=-\rho $, fluctuations of the inflaton are not coupled to fluctuations 
of the perturbed metric. Coupling occurs only as a result of the violation 
of the condition $p=-\rho $.  
\par
Observable quantities can be computed when the initial power spectra 
are known. These are defined in terms of the two-point correlation 
functions. For the Bardeen potential one has
\begin{eqnarray}
\label{defpssca}
\langle 0\vert \Phi _{\rm B}(\eta ,{\bf x})&&\Phi _{\rm B}(\eta ,{\bf x}+{\bf r})
\vert 0\rangle \nonumber \\
& & \equiv \int _0^{+\infty }\frac{{\rm d}n}{n}
\frac{\sin nr}{nr}n^3P_{\rm \Phi _{\rm B}}(\eta ,n),
\end{eqnarray}
whereas for gravitational waves the correlator is given by
\begin{equation}
\label{defpsgw}
\langle 0\vert h_{ij}(\eta ,{\bf x})h^{ij}(\eta ,{\bf x}+{\bf r})
\vert 0\rangle \equiv \int _0^{+\infty }\frac{{\rm d}n}{n}
\frac{\sin nr}{nr}n^3P_{\rm h}(\eta ,n),
\end{equation}
where we have written $h_{ij}=h_{\rm gw}Q_{ij}$. We are specially 
interested in modes which are outside the horizon 
at the end of inflation, i.e. $n/(aH)\ll 1$. For these modes, the power 
spectra do not depend on time and can be written as
\begin{equation}
\label{powespec}
n^3P_{\rm \Phi _{\rm B} }(n)=A_{\rm S}n^{n_{\rm S}-1}, \quad 
n^3P_{\rm h}(n)=A_{\rm T}n^{n_{\rm T}}.
\end{equation}
\par
Let us now consider power law inflation models where the scale factor 
is given by $a(\eta )=l_0\vert \eta \vert ^{1+\beta }$ where $\beta $ 
is a number such that $\beta \le -2$ and $l_0$ has the dimension 
of a length. The advantage of this class of models is that everything 
can be calculated exactly. In the case $\beta =-2$ which corresponds to 
exponential expansion, the length $l_0$ is nothing but the Hubble 
radius, $l_{\rm H}\equiv a^2/a'$ . The function $\gamma $ is a constant given 
by $\gamma =(\beta +2)/(\beta +1)$ which vanishes for $\beta =-2$. We see 
that Eq. (\ref{eomscalar}) now reduces to Eq. (\ref{eomtensor}). The 
spectral indices can be determined exactly and read
\begin{equation}
\label{specindex}
n_{\rm S}=2\beta +5, \quad n_{\rm T}=2\beta +4.
\end{equation}
We have the relation $n_{\rm S}-1=n_{\rm T}$ which is valid exactly only 
for power law inflation. 
\par
Let us now consider a massless scalar field $\Phi (\eta ,{\bf x})$ living in 
a FLRW spacetime. It is convenient to Fourier decompose the field and to 
introduce the quantity $\mu $ defined according to $\Phi (\eta ,{\bf x})
\equiv [1/(2\pi )^{3/2}]\int {\rm d}{\bf n}(\mu /a)e^{i{\bf n}\cdot {\bf x}}$. 
It is easy to show that the Klein-Gordon equation 
reduces to the following equation for $\mu $
\begin{equation}
\label{eomsfield}
\mu ''+\biggl[n^2-\frac{a''}{a}\biggr]\mu =0.
\end{equation} 
This equation is exactly the same as Eq. (\ref{eomtensor}) and 
Eq. (\ref{eomscalar}). Therefore, investigating the properties of 
cosmological perturbations 
is equivalent to investigating the properties of a fictitious scalar 
field $\Phi (\eta ,{\bf x})$. In particular, the calculation of the 
power spectrum of the scalar and tensor perturbations reduces to the 
computation of the power spectrum of this fictitious scalar field. In the 
following, we will restrict our considerations to this case, having 
in mind that, in fact, we will calculate the power spectra of cosmological 
perturbations. 
\par
Let us make a last remark. Although it seems that we have considered 
only a limited class of models (i.e. power law inflation), the previous 
analogy is in fact much more general. This is because the slow roll 
approximation, valid for a wide class of inflationary models, reduces 
to first order to power law inflation.  

\section{Time dependent dispersion relations}

In this section, we present the two classes of modified dispersion relations 
that will be used in this article. Let us return to the 
equation of motion (\ref{eomsfield}). In this equation, the 
presence of the term $n^2$ is due to the differential 
operator $\delta ^{ij}{\rm \partial }_i{\rm \partial }_j$ in the 
Klein-Gordon equation. In Fourier space, this means that
\begin{equation}
\label{normaldr}
\omega ^2=k^2=\frac{n^2}{a^2}.
\end{equation}
The dispersion relation is therefore linear in the physical wavenumber 
$k$: $\omega =k$. A possible alteration of the high frequency 
behaviour of the Klein-Gordon equation can be obtained if we require 
the presence of a nonlinear function $F(k)$ such that $\omega =F(k)$ 
which, for physical wavenumbers smaller than a new characteristic scale 
$k_{\rm C}$, i.e. $k\ll k_{\rm C}$, reduces 
to $\omega \approx k$. This means that the $n^2$ term in the Klein-Gordon 
equation should now be replaced with a time dependent $n_{\rm eff}^2(\eta )$ 
such that
\begin{equation}
\label{neff}
n_{\rm eff}^2=a^2(\eta )F^2(k)=a^2(\eta )F^2[n/a(\eta)].
\end{equation}
We see that, in terms of comoving wavenumbers, we obtain a time 
dependent dispersion relation. In what follows, we will consider 
two explicit examples 
for the function $n_{\rm eff}$. Given the modified dispersion relation, 
Eq. (\ref{eomsfield}) can now be written as 
\begin{equation}
\label{eqommu}
\mu ''+\biggl[n_{\rm eff}^2
-\frac{a''}{a}\biggr]\mu =0.
\end{equation}
Let us analyze this equation in more detail. We can distinguish three 
regimes. In Region I, the wavelength of a given mode, $\lambda (\eta )
\equiv (2\pi /n)a(\eta )$, is much smaller than the characteristic 
length: $\lambda \ll l_{\rm C}$. The nonlinearities in the dispersion 
relation play an important role and the solution of the equation of motion 
depends on the particular form of $F(k)$. A crucial issue is that the mode 
no longer behaves as a free wave initially. As a consequence, the 
choice of initial conditions cannot be done in the usual way. In 
Region II, the wavelength 
of the mode is larger than the characteristic length but still smaller 
than the Hubble radius, $l_{\rm C}\ll \lambda \ll l_{\rm H}$. In this 
case, one can consider the dispersion relation to be linear, i.e. 
$\Omega (\eta )\approx 0$ and neglect the term $a''/a$. Therefore, 
the solution can be expressed as: 
\begin{equation}
\mu _{\rm II}(\eta ) \, = \, B_1e^{in\eta } + B_2e^{-in\eta } \, . 
\end{equation}
Finally, in Region III, the mode is outside the 
Hubble radius: $\lambda \gg l_{\rm H}$ and the solution (the growing mode) is 
given by:
\begin{equation}
\mu _{\rm III}(\eta ) \, = \, Ca(\eta ) \, ,
\end{equation} 
where $C$ is a $n$ dependent 
constant. This constant has to be determined by performing the matching 
of $\mu $ and $\mu '$ at the times of transition between regions I and II 
and regions II and III, $\eta _1$ and $\eta _2$ respectively. Then, the 
spectrum can be calculated and reads
\begin{equation}
\label{spec}
n^3P_{\rm \Phi }=n^3\biggl \vert \frac{\mu }{a}\biggr \vert ^2=n^3\vert C\vert ^2.
\end{equation}
Let us now turn to the first example of a time dependent modified 
dispersion relation.

\subsection{Unruh's dispersion relation}

The dispersion relation used by Unruh in Ref. \cite{Unruh}, in the context of 
black holes physics, is:
\begin{equation}
\label{Udisp}
\omega =F(k)\equiv k_{\rm C}\tanh ^{1/p}\biggl
[\biggl(\frac{k}{k_{\rm C}}\biggr)^p\biggl],
\end{equation}
where $p$ is an arbitrary coefficient. For large values of the 
wave number, this becomes a constant $k_{\rm C}$ 
whereas for small values this is a linear law as expected.
According to Eq. (\ref{neff}), in the context of cosmology, we take
\begin{equation}
\label{Ucos}
n_{\rm eff}(\eta )=\frac{2\pi a(\eta )}{l_{\rm C}}
\tanh ^{1/p}\biggl
[\biggl(\frac{nl_{\rm C}}{2\pi a(\eta )}\biggr)^p\biggl],
\end{equation}
where $l_{\rm C}$ is the 
characteristic length corresponding to $k_{\rm C}$. The 
argument of the hyperbolic tangent can also be 
rewritten as $l_{\rm C}/\lambda (\eta )$. This means that when 
$\lambda \gg l_{\rm C}$, $n_{\rm eff}(\eta )$ tends to $n$.

\subsection{Generalized Corley/Jacobson dispersion relation}

The dispersion relation utilized by Corley and Jacobson in Ref. \cite{CJ} is 
given by the following expression
\begin{equation}
\label{Jdisp}
\omega ^2=F^2(k)\equiv k^2-\frac{k^4}{k_{\rm C}^2}.
\end{equation}
In this article, we consider a more general case and write
\begin{equation}
\label{Jgene}
\omega ^2 = k^2+k^2\sum _{q=1}^{m}b_q\biggl(\frac{k}{k_{\rm C}}\biggr)^{2q},
\end{equation}
where the $b_q$ are, {\it a priori}, arbitrary coefficients. 
Let us suppose that the previous sum only 
contains the last term. The physics depends on the sign of $b_m$. If $b_m$
is negative, then  $\omega $ vanishes for 
$k = k_{\rm C}\vert b_{m}\vert ^{-2m}$. Beyond this point, the dispersion
relation becomes complex. The Corley/Jacobson case corresponds to $m=1$ 
and $b_{1}=-1$. In the context 
of cosmology, the previous ansatz gives rise to the following function 
$n_{\rm eff}(\eta )$
\begin{equation}
\label{defOmJ}
n_{\rm eff}^2(\eta ) = n^2+n^2
\sum _{q=1}^{m}\frac{b_q}{(2\pi )^{2q}}
\biggl(\frac{l_{\rm C}}{a}\biggr)^{2q}n^{2q}.
\end{equation}
Again, when $\lambda \gg l_{\rm C}$ then the effective comoving 
wavenumber simply reduces to $n$. On the other hand, when 
$\lambda \ll l_{\rm C}$, one has
\begin{equation}
n_{\rm eff}^2 \, \approx \,
\frac{b_{m}}{(2\pi )^{2m}}\biggl(\frac{l_{\rm C}}{a}\biggr)^{2m}n^{2m+2} \, .
\end{equation}

\begin{figure}
\begin{center}
\leavevmode
\hspace*{-2.1cm}
\epsfxsize=11cm
\epsffile{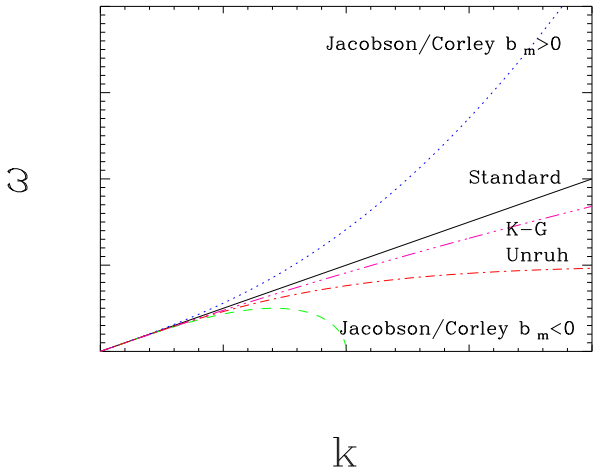}
\end{center}
\caption{Sketch of the different dispersion relations.}
\label{disp}
\end{figure}

The different dispersion relations used in this article are 
displayed in Fig. (\ref{disp}) together with the dispersion 
relation considered in Ref.~\cite{KG} denoted ``KG''.

\section{Quantization of a massive scalar field}

The aim of this section is to develop a Lagrangian and Hamiltonian 
formalism for the system described above. We will show that considering 
a time-dependent dispersion relation is equivalent to giving a 
time-dependent mass to the fictitious scalar field. The main purpose 
of this section is to discuss the initial conditions. As already 
mentioned previously, when the wavelength of a mode is smaller than 
the critical length $l_{\rm C}$, the mode 
does not behave as a free wave because the dispersion relation in this region
is no longer $\omega =k$. As a consequence, it is no longer 
possible to impose the usual initial condition at $\eta =\eta _{\rm i}$, i.e. 
$\mu \rightarrow e^{-in(\eta -\eta _{\rm i})}/\sqrt{2n}$. Another method 
must be used. Following Ref. \cite{Brown78}, we will choose the state which 
initially minimizes the energy density of the field. 

\subsection{Lagrangian and Hamiltonian formalisms}

We now study a massive fictitious scalar field $\Phi$ whose action is given by
\begin{eqnarray}
\label{acfousf}
S &=& \int {\rm d}\eta \int _{R^{3+}}{\rm d}{\bf n}\biggl[
\mu _{\bf n}'\mu _{\bf n}^{*'}+\frac{a^{'2}}{a^2}\mu _{\bf n}\mu _{\bf n}^{*}
\nonumber \\
& &-\frac{a'}{a}(\mu _{\bf n}'\mu _{\bf n}^{*}+\mu _{\bf n}\mu _{\bf n}^{*'})
-n_{\rm eff}^2\mu _{\bf n}\mu _{\bf n}^{*}\biggr].
\end{eqnarray}
In this equation, the scalar field has been Fourier expanded according to 
\begin{equation}
\label{fouphi}
\Phi (\eta ,{\bf x})
=\frac{1}{(2\pi )^{3/2}}\frac{1}{a(\eta )}\int {\rm d}{\bf n}\mu _{\bf n}(\eta )
e^{i{\bf n}\cdot {\bf x}},
\end{equation}
and $\mu _{\bf n}(\eta )$ denotes the complex Fourier component of 
the field. We can easily check that the Lagrange equation of motion for the 
quantity $\mu _{\bf n}(\eta )$ leads to Eq. (\ref{eqommu}).
\par
We are now in a position where we can pass to the Hamiltonian 
formalism. Our first move is to perform the following time-dependent 
transformation
\begin{equation}
\label{trans1}
\mu _{\bf n}(\eta )\equiv \frac{1}{N(n,\eta )}\psi _{\bf n}(\eta ),
\end{equation}
where $N(n,\eta )$ is a time-dependent factor which will be fixed 
below. Next, the action given in Eq. (\ref{acfousf}) expressed in terms 
of the new variable $\psi _{\bf n}(\eta )$ takes the form
\begin{eqnarray}
\label{acfousf2}
S &=& \int {\rm d}\eta \int _{R^{3+}}{\rm d}{\bf n}\biggl[
\frac{1}{N^2}\psi _{\bf n}'\psi _{\bf n}^{*'} 
+\frac{1}{N^2}\biggl(\frac{N'}{N}+\frac{a'}{a}\biggr)^2\psi _{\bf n}\psi _{\bf n}^{*}
\nonumber \\
& &-\frac{1}{N^2}\biggl(\frac{N'}{N}+\frac{a'}{a}\biggr)
\biggl(\psi _{\bf n}'\psi _{\bf n}^{*}+\psi _{\bf n}
\psi _{\bf n}^{*'}\biggr) \nonumber \\
& &-\frac{1}{N^2}n_{\rm eff}^2
\psi  _{\bf n}\psi _{\bf n}^{*}\biggr].
\end{eqnarray}
We can now calculate the conjugate momentum to $\psi _{\bf n}(\eta )$. Its 
definition is $p_{\bf n}
\equiv {\rm \partial }\bar{{\cal L}}_{\bf n}/{\rm \partial }
\psi _{\bf n}^*(\eta )$ where $\bar{{\cal L}}_{\bf n}$ is the Lagrangian 
density (the bar indicates that one calculates the Lagrangian in 
Fourier space) which one can deduce from the previous equation. The conjugate 
momentum reads
\begin{equation}
\label{conmom}
p _{\bf n}=\frac{1}{N^2}\biggl(\psi _{\bf n}'-\frac{a'}{a}\psi _{\bf n}\biggr)
-\frac{N'}{N^3}\psi _{\bf n}.
\end{equation}
The Hamiltonian can be determined using the following relation
\begin{equation}
\label{defH}
\bar{{\cal H}}_{\bf n}\equiv p _{\bf n}\psi _{\bf n}^{*'}
+p_{\bf n}^*\psi _{\bf n}'-\bar{{\cal L}}_{\bf n}.
\end{equation}
Inserting the expressions of the Lagrangian and of the the conjugate momentum 
in this definition, we obtain
\begin{eqnarray}
\label{expbarH}
\bar{{\cal H}}_{\bf n} &=& N^2p _{\bf n}p_{\bf n}^*+\frac{(aN)'}{aN}\biggl(
\psi _{\bf n}p_{\bf n}^*+\psi _{\bf n}^{*}p _{\bf n}\biggr) \nonumber \\
& &+\frac{1}{N^2}n_{\rm eff}^2 \psi  _{\bf n}\psi _{\bf n}^{*}.
\end{eqnarray}
The explicit quantization can now be carried out. We express the Fourier component 
$\psi _{\bf n}$ and its conjugate momentum $p_{\bf n}$ in terms of creation 
and annihilation operators, satisfying the usual commutation relation 
$[c_{\bf n},c_{\bf r}^{\dagger }]=\delta ({\bf n}-{\bf r})$, according to
\begin{equation}
\label{cre12}
\psi _{\bf n} \equiv \sqrt{\hbar }\biggl(c_{\bf n}+
c_{-{\bf n}}^{\dagger }\biggr), \quad 
p_{\bf n} \equiv \frac{\sqrt{\hbar}}{2i}
\biggl(c_{\bf n}-c_{-{\bf n}}^{\dagger }\biggr).
\end{equation}
The Hamiltonian operator is obtained by plugging the previous expressions
into Eq. (\ref{expbarH}) and requiring that `$\hbar \omega /2$' be present 
in each mode, which fixes the normalization factor $N$ to be 
\begin{equation}
N^2 \, = \, 2\omega (\eta ) \, ,
\end{equation}
where 
$\omega $ is the `comoving frequency' defined 
by $\omega (\eta )\equiv n_{\rm eff}$. Although we 
use the same notation for convenience, this frequency should not be confused 
with the physical frequency which appears in Eqns. (\ref{Udisp}) and (\ref{Jdisp}) 
and which can obtained by multiplying the comoving frequency by a factor 
$1/a$. The Hamiltonian reads
\begin{eqnarray}
\label{expH}
H = & &\int _{R^3}{\rm d}{\bf n}\biggl[\frac{\hbar\omega }{2}
\biggl(c_{\bf n}c_{\bf n}^{\dagger }
+c_{-{\bf n}}c_{-{\bf n}}^{\dagger }\biggr)
+\frac{i\hbar }{2}\frac{(a\sqrt{\omega } )'}{a\sqrt{\omega }}
\biggl(c_{-{\bf n}}^{\dagger}c_{\bf n}^{\dagger } \nonumber \\
& & -c_{-{\bf n}}c_{\bf n}\biggr)\biggr].
\end{eqnarray} 
This Hamiltonian has the usual structure. The first term is just a 
collection of harmonic oscillators whereas the second term 
represents the interaction between the background and the perturbations. This 
term is responsible for the phenomenon of particle creation, which is a 
squeezing effect. In a static spacetime, the pump 
function $(a\sqrt{\omega })'/(a\sqrt{\omega })$ vanishes and the interaction 
part of the Hamiltonian disappears. The field operator can be 
expressed as
\begin{eqnarray}
\label{quantsf}
\Phi (\eta ,{\bf x}) &=&
\frac{\sqrt{\hbar }}{a(\eta )}\frac{1}{(2\pi )^{3/2}}
\int \frac{{\rm d}{\bf n}}{\sqrt{2\omega (\eta )}}\biggl[
c_{\bf n}(\eta )e^{i{\bf n}\cdot {\bf x}} \nonumber \\
& &+ c_{\bf n}^{\dagger }(\eta )e^{-i{\bf n}\cdot {\bf x}}\biggr].
\end{eqnarray}
The time evolution of the creation and annihilation operators and therefore 
of the quantum scalar field is calculated by means of the 
Heisenberg equation:
\begin{equation}
\label{Heieq}
i\hbar \frac{{\rm d}}{{\rm d}\eta }c_{\bf n}(\eta )=[c_{\bf n},H].
\end{equation}
Using the form of the Hamiltonian derived previously, one gets the following 
equations of motion
\begin{eqnarray}
\label{Heisen1}
i\hbar \frac{{\rm d}c_{\bf n}}{{\rm d}\eta } &=& \hbar \omega (\eta )c_{\bf n}
+i\hbar \frac{(a\sqrt{\omega })'}{a\sqrt{\omega }}c_{-{\bf n}}^{\dagger }, \\
\label{Heisen2}
i\hbar \frac{{\rm d}c_{\bf n}^{\dagger }}{{\rm d}\eta } &=& 
-\hbar \omega (\eta )c_{\bf n}^{\dagger }
+i\hbar \frac{(a\sqrt{\omega })'}{a\sqrt{\omega }}c_{-{\bf n}}.
\end{eqnarray}
The solution of these equations is a Bogoliubov transformation which can be 
written as
\begin{eqnarray}
\label{Bog1}
c_{\bf n}(\eta ) &=& u_n(\eta )c_{\bf n}(\eta _{\rm i})+
v_n(\eta )c_{-{\bf n}}^{\dagger }(\eta _{\rm i}), \\
\label{Bog2}
c_{\bf n}^{\dagger }(\eta ) &=& u_n^*(\eta )c_{\bf n}^{\dagger }(\eta _{\rm i})
+v_n^*(\eta )c_{-{\bf n}}(\eta _{\rm i}),
\end{eqnarray}
where we have introduced two new functions $u_n(\eta )$ and $v_n(\eta )$. These 
functions satisfy $\vert u_n(\eta )\vert ^2-\vert v_n(\eta )\vert ^2=1$ 
in order for the commutation relation given to be preserved in time. Let us notice 
that $u_n$ and $v_n$ do not depend on the vector ${\bf n}$ but only 
on its modulus $n$. Inserting the 
previous equations in Eqns. (\ref{Heisen1}) and (\ref{Heisen2}), one obtains 
the equation of motion for these two functions
\begin{eqnarray}
\label{eomu}
i\hbar \frac{{\rm d}u_n}{{\rm d}\eta } &=& \hbar \omega (\eta )u_n
+i\hbar \frac{(a\sqrt{\omega })'}{a\sqrt{\omega }}v_n^*, \\
\label{eomv}
i\hbar \frac{{\rm d}v_n}{{\rm d}\eta } &=& 
\hbar \omega (\eta )v_n
+i\hbar \frac{(a\sqrt{\omega })'}{a\sqrt{\omega }}u_n^*.
\end{eqnarray}
The functions $u_n$ and $v_n$ can be re-expressed in terms of three 
other arbitrary functions $r_n(\eta )$, $\theta _n(\eta )$ and 
$\varphi _n(\eta )$. Following this path would lead to the squeezed state 
formalism. However, we will not need it in this article.

\subsection{Fixing the initial conditions}

The previous considerations permit to fix the initial 
value of the mode function $\mu _n(\eta _{\rm i})$ and its 
derivative $\mu _n'(\eta _{\rm i})$ for any choice of 
function $\Omega (\eta )$, i.e. for any time dependent 
dispersion relation.
\par
It is straightforward to check that the function
\begin{equation}
\label{linkuvmu}
\mu _n\equiv \frac{1}{N(n,\eta )}(u_{n}+v_{n}^*)=
\frac{1}{\sqrt{2 \omega }}(u_{n}+v_{n}^*),
\end{equation}
satisfies Eq. (\ref{eqommu})\footnote{It should be noticed that 
$\mu _n$ is not exactly the mode function introduced before. It 
is dimensionless (instead of dimension $\sqrt{\hbar c}$) and 
depends only on the modulus $n$. In the same manner, we now 
deal with a `new' function $\psi _n\equiv N\mu _n$.}. From Eqns. (\ref{Bog1}) 
and (\ref{Bog2}), we see 
that the initial conditions for the two function $u_n$ and $v_n$ are given 
by: $u_{n}(\eta =\eta _{\rm i})=1$ and 
$v_{n}(\eta =\eta _{\rm i})=0$. Therefore, the initial value of 
the mode function $\mu $ can be written as:
\begin{equation}
\label{inimu}
\mu (\eta =\eta _{\rm i})=\frac{1}{\sqrt{2\omega (\eta _{\rm i})}}=
\frac{1}{\sqrt{2n_{\rm eff}}}.
\end{equation}
Let us now turn to the determination of $\mu '(\eta =\eta _{\rm i})$. It 
will be found by the requirement that the energy density is minimized.
The stress energy tensor can be obtained from the action (\ref{acfousf}) 
with the help of the standard definition. In terms of the Fourier 
components $\psi _{n}$, the energy density reads
\begin{eqnarray}
\label{exprho}
\rho &=& \frac{\hbar }{4\pi ^2a^4}\int _0^{\infty }
\frac{{\rm d}n}{N^2}\biggl[
\psi _{n}'\psi _{n}^{*'}-\frac{(aN)'}{aN}( 
\psi _{n}\psi _{n}^{*'} 
+\psi _{n}'\psi _{n}^{*}) \nonumber \\
& & +\frac{a'^2}{a^2}\psi _{n}\psi _{n}^{*}
+\frac{N'^2}{N^2}\psi _{n}\psi _{n}^{*}
+n_{\rm eff}^2\psi _{n}\psi _{n}^{*}
\nonumber \\
& & +2\frac{a'N'}{aN}\psi _{n}\psi _{n}^{*}\biggr].
\end{eqnarray}
We now define the functions $x(\eta ) $ and $y(\eta )$ as the real 
and imaginary parts of the ratio $\psi _{n}'/\psi _{n} \equiv 
x+iy$, respectively. Then, the initial energy density can be 
expressed in terms of $x_{\rm i}\equiv x(\eta =\eta _{\rm i})$, 
$y_{\rm i}\equiv y(\eta =\eta _{\rm i})$ and the Wronskian 
$W(n)\equiv \mu _n'\mu _n^*-\mu _n^{*'}\mu _n$ which is a time independent 
quantity (as can be checked in calculating ${\rm d}W(n)/{\rm d}\eta $ and 
using the equation of motion for $\mu _n$)
\begin{eqnarray}
\label{rhoini}
\rho &=& \frac{\hbar}{4\pi ^2a^4}\int _0^{\infty }{\rm d}n
\frac{W(n)}{2iy_{\rm i}}\biggl[x_{\rm i}^2+y_{\rm i}^2-2\frac{(aN)'}{aN}x_{\rm i}
+\frac{a'^2}{a^2} \nonumber \\
&+& n_{\rm eff}^2+\frac{N'^2}{N^2}
+2\frac{a'N'}{aN}\biggr],
\end{eqnarray}
where $N$ and $a$ are also evaluated at the initial time. Notice that, while 
deriving the previous equation, we used the fact that the Wronskians of $\mu _n$ 
and $\psi _n$ are related by a factor $N^2$. The `vacuum'  used 
in this article is defined as the state which initially minimizes the energy 
density. The variation of the previous expression with respect to $x_{\rm i}$ 
and $y_{\rm i}$ leads to
\begin{eqnarray}
\label{varrho}
\delta \rho &=& \frac{\hbar}{4\pi ^2a^4}
\int _0^{\infty }{\rm d}n
\frac{W(n)}{2i}\biggl\{\frac{2}{y_{\rm i}}\biggl[x_{\rm i}-\frac{(aN)'}{aN}\biggr]
\delta x_{\rm i} \nonumber \\
& &+\frac{1}{y_{\rm i}^2}\biggl[y_{\rm i}^2-x_{\rm i}^2+2\frac{(aN)'}{aN}x_{\rm i}
-\frac{a'^2}{a^2}-n_{\rm eff}^2 \nonumber \\
& &- \frac{N'^2}{N^2}
-2\frac{a'N'}{aN}\biggl]\delta y_{\rm i}\biggr\}.
\end{eqnarray}
Demanding that $\delta \rho =0$, one deduces the initial values of 
$x$ and $y$
\begin{equation}
\label{inixy}
x_{\rm i}=\frac{a'}{a}(\eta _{\rm i})+\frac{N'}{N}(\eta _{\rm i}), \quad 
y_{\rm i}=\pm n_{\rm eff}.
\end{equation}
These expressions can be simplified. Using the explicit form 
of the function $N(n,\eta )$, one can write
\begin{equation}
\label{NN'}
\frac{N'}{N}=\frac{\omega '}{2 \omega}.
\end{equation}
At the time $\eta =\eta _{\rm i}$, it is reasonable to consider 
$\lambda \ll l_{\rm C}$ (otherwise, the whole problem studied here 
would be pointless). Then, for Unruh's dispersion relation, one 
finds $N'/N\approx a'/2a$ and for the Corley/Jacobson dispersion relation, one 
has $N'/N\approx -ma'/2a$. In addition, $a'/a$ is very small in the limit 
where the conformal time goes to $-\infty $ since $a'/a(\eta _{\rm i})=
(1+\beta )/\vert \eta _{\rm i}\vert $ and $\vert \eta _{\rm i}\vert \gg 1$. 
Therefore, one gets that 
$\psi _{n}'/\psi _{n}\approx iy_{\rm i}$. On the other 
hand, we have $\mu _n=\psi _n/N$. Combining this formula with the 
previous one, one obtains $\mu _n'+N'/N=iy_{\rm i}\mu _n$. Neglecting 
again the term $N'/N$, we finally arrive at
\begin{equation}
\label{inideriv}
\mu '(\eta =\eta _{\rm i})=\pm i\sqrt{\frac{n_{\rm eff}}{2}}.
\end{equation}
The initial conditions are now completely fixed and given 
by Eqns. (\ref{inimu}) and (\ref{inideriv}). 
\par
Let us also mention that it is possible to adopt another choice of 
initial conditions which corresponds to 
the `instantaneous Minkowski vacuum' at 
$\eta =\eta _{\rm i}$, namely 
\begin{equation}
\label{Minvac}
\mu (\eta _{\rm i})=\frac{1}{\sqrt{2n}}, \quad \mu ' (\eta _{\rm i})=
\pm i\sqrt{\frac{n}{2}}.
\end{equation}
If the dispersion relation is standard, then $\Omega =0$ and the mode 
is initially free: locally, it does not feel the curvature of space-time 
and behaves as it were flat. In this case, the two possible choices of 
initial conditions discussed above coincide.
\par
Two last comments are in order before ending this section. Let us 
first remark that the concept of an initial state which 
minimizes the energy density of the field could be problematic in 
a region where the dispersion relation becomes complex, as it 
is the case for the Corley/Jacobson dispersion relation with 
$b_m<0$, since the energy needs not to be bounded from below in 
such a situation. We are not aware of any more obvious method 
than the one used here to deal with this case.
\par
Finally, although we have introduced two initial states, it should be 
clear that the minimizing energy state is the only physical vacuum 
state. The instantaneous Minkowski vacuum is considered here only to 
stress the fact that the choice of the initial conditions becomes more 
crucial than in the standard situation where one can show that a large 
class of initial states leads to the same spectrum \cite{BH85} 
(although, as already mentioned in the introduction, it is possible to 
find examples which do not belong to this class of initial 
states \cite{MRS99}).

\section{Analytical solutions}

\begin{figure}
\begin{center}
\leavevmode
\hspace*{-2cm}
\epsfxsize=9cm
\epsffile{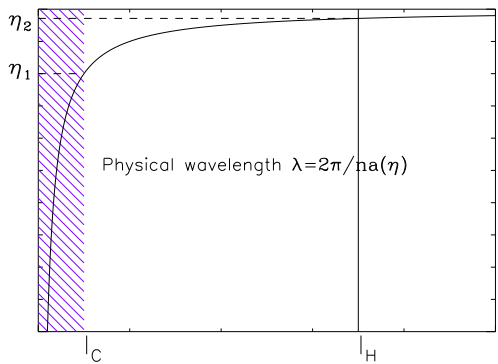}
\end{center}
\caption{Sketch of the evolution of a physical mode throughout the 
three regions defined in the text. The dashed region is the region 
where the dispersion relation is modified.}
\label{sketch}
\end{figure}

In this section, we calculate the spectrum of fluctuations for the two 
classes of dispersion relations introduced in Section III. We focus 
on a fixed comoving wavenumber $n$ and proceed as follows. We 
solve the equation of motion in each of the three regions (defined 
in Section III) separately. The coefficients of the two fundamental 
solutions in Region I are fixed by the initial conditions discussed above. Then, we 
explicitly perform the matching of $\mu $ and $\mu '$ at the 
transitions between Region I 
and Region II, which occurs at a time denoted by $\eta _1$, and between 
Region II and Region III, which occurs at time $\eta _2$, to obtain the 
coefficients of the two fundamental solutions in Region III, from which 
the spectrum can be calculated. The time $\eta _2$ is when the mode
crosses the Hubble radius, which is given by 
\begin{equation} 
\label{HR}
l_{\rm H} (\eta )\, = \, \frac{l_0}{\vert 1+\beta \vert }\vert \eta \vert ^{2+\beta } \, .
\end{equation}
Thus, the condition  $l_{\rm H}(\eta _2)=\lambda (\eta _2)$ boils 
down to 
\begin{equation}
\label{time2}
\vert \eta _2\vert =\frac{2\pi}{n} \vert 1+\beta \vert \, .
\end{equation} 
The geometry of space-time is illustrated in Fig. (\ref{sketch}).

We start this section with Unruh's dispersion relation.

\subsection{Unruh's case}

The equation of motion for the mode function can be written as
\begin{equation}
\label{eomU}
\mu ''+\biggl\{\frac{4\pi ^2}{l_{\rm C}^2}a^2
\tanh ^{2/p}\biggl
[\biggl(\frac{l_{\rm C}}{\lambda (\eta )}\biggr)^p\biggl]
-\frac{a''}{a}\biggr\}\mu =0.
\end{equation}
This equation can be solved exactly in Region I only if the scale 
factor is given by $a(\eta )=l_0/\vert \eta \vert$, i.e. 
in the case $\beta =-2$. Fortunately, this corresponds to the 
de Sitter space-time, the prototypical model of inflationary 
cosmology. Note that in this case $l_0$ is the Hubble 
radius [see Eq. (\ref{HR})]. In 
Region I, the hyperbolic tangent is approximatively 
one since $l_{\rm C}\gg \lambda $ initially. Therefore, 
Equation (\ref{eomU}) reduces to  
\begin{equation}
\label{eomUI}
\mu ''+\biggl(\frac{4\pi ^2l_0^2/l_{\rm C}^2-2}{\eta ^2} \biggr)\mu =0.
\end{equation}
Note that, in fact, the form of this last equation is independent 
of the precise form (i.e. the hyperbolic tangent) of the 
dispersion relation in the regime $l_{\rm C}\gg \lambda $. It 
is just necessary to assume that $F(k)$ goes to a constant. 
We see that the 
result depends in an essential way on the dimensionless parameter 
$\epsilon \equiv l_{\rm C}/l_0$. At this point, we have assumed nothing 
about the value of the ratio $l_{\rm C}/l_0$. However, physically, it is 
clear that $\epsilon \ll 1$. One would expect the cutoff length to 
be given by the Planck length ($l_{\rm C}\approx l_{\rm Pl}$), whereas
$l_0\approx 10^5l_{\rm Pl}$ if the spectrum of fluctuations is COBE 
normalized. In this case, we have $\epsilon \approx 10^{-5}$. In the 
following, we will use an expansion in terms of this parameter. The 
exact solution of Eq. (\ref{eomUI}) is
\begin{equation}
\label{solU}
\mu _{\rm I}(\eta )=A_1\vert \eta \vert ^{x_1}+A_2\vert \eta \vert ^{x_2},
\end{equation}
where the exponents $x_1$ and $x_2$ are given by
\begin{equation}
\label{expU}
x_{1,2}=\frac{1}{2}\pm \frac{1}{2}\sqrt{9-\frac{16\pi ^2}{\epsilon ^2}}.
\end{equation}
\par
It is now time to fix the coefficients $A_1$ and $A_2$. They are  
completely determined by the initial conditions (\ref{inimu}) 
and (\ref{inideriv}). In the approximation where $l_{\rm C}\gg \lambda $, 
they are solutions of (note that we do not yet use the fact that $\epsilon $ 
is small)
\begin{eqnarray}
\label{iniU1}
A_1\vert \eta _{\rm i}\vert ^{x_1}+A_2\vert \eta _{\rm i}\vert ^{x_2} &=& 
\frac{1}{2}\sqrt{\frac{\epsilon }{\pi }}\vert \eta _{\rm i}\vert ^{1/2}, \\
\label{iniU2}
A_1x_1\vert \eta _{\rm i}\vert ^{x_1-1}+A_2x_2\vert \eta _{\rm i}\vert ^{x_2-1} &=& 
\mp i\sqrt{\frac{\pi }{ \epsilon }}\vert \eta _{\rm i}\vert ^{-1/2}.
\end{eqnarray}
The exact solution of this system of equations can be written as
\begin{eqnarray}
\label{soluU1}
A_1 &=& \vert \eta _{\rm i}\vert ^{1/2-x_1}\frac{1}{1-x_1/x_2}
\frac{1}{2}\sqrt{\frac{\epsilon}{\pi }}\biggl(
1\pm \frac{2i\pi }{\epsilon x_2}\biggr), \\
\label{soluU2}
A_2 &=& \vert \eta _{\rm i}\vert ^{1/2-x_2}\frac{1}{1-x_2/x_1}
\frac{1}{2}\sqrt{\frac{\epsilon}{\pi }}\biggl(
1\pm \frac{2i\pi }{\epsilon x_1}\biggr).
\end{eqnarray}
It is at this point that we use the fact that $\epsilon$ is small. To 
first order in a systematic expansion in this parameter we obtain
\begin{eqnarray}
\label{solapproU1}
A_1 &\approx &\frac{i}{8}\biggl(\frac{\epsilon }{\pi}\biggr)^{3/2}
\biggl(\frac{1}{2}-\frac{2i\pi }{\epsilon}\pm \frac{2i\pi }{\epsilon}\biggr)
\nonumber \\
& & \times \exp \biggl(
-\frac{2i\pi }{\epsilon }\ln \bigl\vert \eta _{\rm i} \bigr\vert \biggr), \\
\label{solapproU2}
A_2 &\approx &\frac{i}{8}\biggl(\frac{\epsilon }{\pi }\biggr)^{3/2}
\biggl(\frac{1}{2}+\frac{2i\pi }{\epsilon}\pm \frac{2i\pi }{\epsilon}\biggr)
\nonumber \\
& & \times \exp \biggl(
\frac{2i\pi }{\epsilon }\ln \bigl\vert \eta _{\rm i} \bigr \vert \biggr).
\end{eqnarray}
We now pursue the calculation for both choices of the sign of the 
initial conditions. We introduce an index `u' for the upper 
choice and `l' for the lower choice. This leads to
\begin{eqnarray}
\label{A1Uu}
A_1^u &=& \frac{i}{16}\biggl(\frac{\epsilon }{\pi}\biggr)^{3/2}
\exp \biggl(
-\frac{2i\pi }{\epsilon }\ln \bigl\vert \eta _{\rm i} \bigr\vert \biggr), \\
\label{A2Uu}
A_2^u &=& \frac{1}{2}\biggl(\frac{\epsilon }{\pi}\biggr)^{1/2}
\exp \biggl(
\frac{2i\pi }{\epsilon }\ln \bigl\vert \eta _{\rm i} \bigr\vert \biggr), \\
\label{A1Ul}
A_1^l &=& \frac{1}{2}\biggl(\frac{\epsilon }{\pi}\biggr)^{1/2}
\exp \biggl(
-\frac{2i\pi }{\epsilon }\ln \bigl\vert \eta _{\rm i} \bigr\vert \biggr), \\
\label{A2Ul}
A_2^l &=& -\frac{i}{16}\biggl(\frac{\epsilon }{\pi}\biggr)^{3/2}
\exp \biggl(
\frac{2i\pi }{\epsilon }\ln \bigl\vert \eta _{\rm i} \bigr\vert \biggr).
\end{eqnarray}
Therefore, one has $A_2^u \gg A_1^u$,  $A_1^l \gg A_2^l$ and only one branch 
of the solution 
(\ref{solU}) survives. Then, the solution in Region I can be expressed as
\begin{equation}
\label{solUI}
\mu _{\rm I}^{u,l}(\eta )= \frac{1}{2}\sqrt{\frac{\epsilon \vert \eta \vert}{\pi}}
\exp \biggl(\mp \frac{2i\pi }{\epsilon }\ln \biggl\vert 
\frac{\eta }{\eta _{\rm i} }\biggr\vert \biggr).
\end{equation}

Let us now turn to Region II. As already mentioned above, in this region, 
the solution is given by 
\begin{equation}
\mu _{\rm II}(\eta ) \, = \, B_1e^{in\eta }+
B_2e^{-in\eta } \, .
\end{equation} 
The coefficients $B_1$ and $B_2$ are determined by the matching 
of this solution with the solution (\ref{solUI}) at the time 
$\eta _1$. Continuity of $\mu $ and $\mu '$ yields
\begin{eqnarray}
\label{eqB1}
inB_1e^{in\eta _1} &+& inB_2e^{-in\eta _1} \nonumber \\ 
&=&\frac{in}{2}
\biggl(\frac{\epsilon \vert \eta _1\vert}{\pi }\biggr)^{1/2}
\exp \biggl(\mp \frac{2i\pi }{\epsilon }\ln \biggl\vert 
\frac{\eta _1}{\eta _{\rm i} }\biggr\vert \biggr), \\
\label{eqB2}
inB_1e^{in\eta _1} &-& inB_2e^{-in\eta _1} \nonumber \\
&=& \pm i\biggl(\frac{\pi}{\epsilon \vert \eta _1\vert}\biggr)^{1/2}  
\exp \biggl(
\mp \frac{2i\pi }{\epsilon }\ln \biggl\vert 
\frac{\eta _1}{\eta _{\rm i }}\biggr \vert \biggr).
\end{eqnarray}
The solution can be found easily and reads
\begin{eqnarray}
\label{Bis1}
B_1 &=& \frac{1}{2\sqrt{n}}(1\pm 1)\exp \biggl(
\mp \frac{2i\pi }{\epsilon }\ln \biggl\vert \frac{\eta _1}{\eta _{\rm i} }\biggr \vert 
+\frac{2i\pi }{\epsilon }\biggr), \\
\label{Bis2}
B_2 &=& \frac{1}{2\sqrt{n}}(1\mp 1)\exp \biggl(
\mp \frac{2i\pi }{\epsilon }\ln \biggl\vert \frac{\eta _1}{\eta _{\rm i} }\biggr \vert 
-\frac{2i\pi }{\epsilon }\biggr), 
\end{eqnarray}
As a consequence, the solution in Region II also contains only one 
branch. 
\par
Finally, we must solve the mode equation in Region III. As 
already mentioned, the non-decaying mode is 
\begin{equation}
\mu _{\rm III}=Ca(\eta ) \, . 
\end{equation}
The coefficient $C$ 
is fixed by the matching of the mode function when the mode crosses the 
horizon at $\eta _2$ \footnote{To be more precise, we should take the 
decaying mode in Region III into account and match both $\mu$ and 
$\mu^{\prime}$ at time $\eta_2$. This only changes the result by 
an unimportant constant of order one.}. One gets 
\begin{equation} C \, = \, \mu _{\rm II}(\eta _2)\frac{\vert 
\eta _2\vert}{l_0}= \frac{2\pi}{n} \frac{\mu _{\rm II}(\eta _2)}{l_0} \, .
\end{equation}
Therefore, regardless of the choice of the sign of the initial 
conditions, we have $\vert C\vert \propto 1/n^{3/2}$ and as a result
\begin{equation}
\label{specU}
n^3P_{\rm \Phi }\propto n^0 \, .
\end{equation}
We see that, when $\beta =-2$, the final answer is not changed 
compared to what is obtained without the modification of the dispersion 
relation, i.e. we get a scale invariant spectrum 
$n_{\rm S}=1$ [see Eq. (\ref{specindex})]. 
\par
We now discuss different initial conditions. We adopt the `instantaneaous 
Minkowski' initial conditions given by Eqns. (\ref{Minvac}). Of course, the 
form of the solution in Region I is still the same but, now, the coefficients 
$A_1$ and $A_2$ are different. The exact expressions for theses coefficients 
can now be written as
\begin{eqnarray}
\label{AM1}
A_1 &=& \pm \sqrt{\frac{n}{2}}\frac{i}{x_2-x_1}
\vert \eta _{\rm i}\vert ^{1-x_1}\biggl(1\mp \frac{ix_2}
{n\vert \eta _{\rm i}\vert }\biggr), \\
\label{AM2}
A_2 &=& \pm \sqrt{\frac{n}{2}}\frac{i}{x_1-x_2}
\vert \eta _{\rm i}\vert ^{1-x_2}\biggl(1\mp \frac{ix_1}
{n\vert \eta _{\rm i}\vert }\biggr).
\end{eqnarray}
In the limit when the parameter $\epsilon $ is small, an expansion 
of the previous expressions leads to the following formulas
\begin{eqnarray}
\label{AapproM1}
A_1 \approx \frac{1}{2\sqrt{2n}}\vert \eta _{\rm i}\vert ^{-2i\pi /\epsilon -1/2}, 
\quad 
A_2 \approx \frac{1}{2\sqrt{2n}}\vert \eta _{\rm i}\vert ^{2i\pi /\epsilon -1/2}. 
\end{eqnarray}
The result does not depend on the choice of the sign of the initial 
conditions. We see also another  crucial difference in 
comparison with the previous case, 
see Eq. (\ref{solapproU1}) and (\ref{solapproU2}): this time, the coefficients 
are of the same order in $\epsilon$. Therefore, the solution in 
Region I is now given by a cosine instead of by a pure phase
\begin{equation}
\label{solIUM}
\mu _{\rm I}(\eta )=\frac{1}{\sqrt{2n}}\biggl\vert 
\frac{\eta }{\eta _{\rm i}}\biggr \vert 
^{1/2}\cos \biggl(\frac{2\pi }{\epsilon }\ln 
\biggl\vert \frac{\eta }{\eta _{\rm i}}\biggr \vert \biggr).
\end{equation}
The solution in Region II is still given by plane waves. The matching 
at time $\eta _1$ permits the calculation of the coefficients $B_1$ and 
$B_2$. They read
\begin{eqnarray}
\label{B1M}
B_1 &=& \frac{1}{2n}\sqrt{\frac{\pi }{\epsilon \vert \eta _{\rm i}\vert }}
\exp \biggl(-in\eta _1-\frac{2\pi i}{\epsilon }\ln 
\biggl \vert \frac{\eta _1}{\eta _{\rm i}}\biggr \vert \biggr), \\
\label{B2M}
B_2 &=& \frac{1}{2n}\sqrt{\frac{\pi }{\epsilon \vert \eta _{\rm i}\vert }}
\exp \biggl(in\eta _1+\frac{2\pi i}{\epsilon }\ln 
\biggl \vert \frac{\eta _1}{\eta _{\rm i}}\biggr \vert \biggr).
\end{eqnarray}
Again, there is an important difference in comparison with the 
previous case: both coefficients are now non vanishing. The 
mode function in Region II can be expressed as  
\begin{equation}
\label{solMII}
\mu _{\rm II}(\eta )=\frac{1}{n}
\sqrt{\frac{\pi }{\epsilon \vert \eta _{\rm i}\vert}}
\cos \biggl(n\eta -n\eta _1 +\frac{2\pi }{\epsilon }\ln 
\biggl \vert \frac{\eta _1}{\eta _{\rm i}}\biggr \vert \biggr).
\end{equation}
The function is proportional to $1/n$ instead of $1/\sqrt{n}$.
The determination of the constant $C$ proceeds as previously and 
leads to the spectrum
\begin{equation}
\label{specUM}
n^3P_{\rm \Phi }\propto n^{-1}\cos ^2\biggl(\frac{2\pi }{\epsilon }+
\frac{2\pi }{\epsilon }\ln \biggl \vert \frac{2\pi }{n\eta _{\rm i}}\biggr 
\vert \biggr).
\end{equation}
A few remarks are in order here. Firstly, the difference 
between (\ref{specU}) and (\ref{specUM}) demonstrates that the 
final result does depend on the choice of the initial conditions. Secondly, 
the spectral index is now modified and is $n_{\rm S}=0$ instead of 
$n_{\rm S}=1$ previously. Thirdly, oscillations in the spectrum are 
present. If $n_1$ and $n_2$ are two wave numbers such that the argument 
of the cosine differs by a factor $2\pi p$ where $p$ is an integer then 
one has $n_2/n_1=\exp (p\epsilon)$. This means that unless $p$ is comparable 
to $\epsilon^{-1} $, $n_1$ and $n_2$ are almost equal. Therefore, the oscillations 
are very rapid.

\subsection{The Corley/Jacobson case}

With the dispersion relation (\ref{Jgene}), the equation of motion becomes
\begin{equation}
\label{eomJ}
\mu ''+\mu \biggl[n^2+n^2\sum _{q=1}^{m}\frac{b_q}{(2\pi )^{2q}}
\biggl(\frac{\epsilon \, n}{\vert \eta \vert ^{1+\beta }}\biggr)^{2q}
-\frac{a''}{a}\biggr]=0.
\end{equation}
This equation is valid for any scale factor of the 
form $a(\eta )=l_0\vert \eta \vert ^{1+\beta }$. Unlike in Unruh's 
case, we do not need to specify $\beta =-2$. 
\par
We now need to discuss the form of the solution in 
Region I. This crucially depends on the sign of the 
coefficient $b_{m}$. In the regime we are interested 
in, i.e. $l_{\rm C}\gg \lambda (\eta _{\rm i})$, one can 
retain only the dominate term and the dispersion relation can be 
written as
\begin{equation}
\label{initialCJ}
n_{\rm eff}^2\approx n^2+n^2b_{m}
\biggl(\frac{l_{\rm C}}{\lambda }\biggr)^{2m}.
\end{equation}
This means that if $b_{m}$ is positive, the dispersion relation 
remains real. If $b_{m}$ is negative the situation is 
more complicated. For very small value of $|b_{m}|$, the dispersion 
relation can remain real even in the regime $l_{\rm C}\gg 
\lambda (\eta _{\rm i})$. However, it seems a bit artificial 
to fine-tune the value of $|b_m|$ such that this actually 
happens. Without this fine-tuning the dispersion relation certainly 
becomes complex. This last property should not be considered 
as a surprise. Indeed there exist many situations in Physics 
where complex dispersion relations appear. This is for example 
the case in Hydrodynamics when one describes the damping of 
a sound wave in a fluid due to viscosity \cite{LL}. Then, the 
dispersion relation is given by $k=\omega /c+ia\omega ^2$ where $a$ 
is a factor which depends on the viscosity coefficients. In 
Cosmology, other examples are Silk damping or damping of density 
perturbations due to neutrino decoupling \cite{SSW}. In this paper, 
we choose to analyze both cases and write 
$b_m\equiv s\vert b_m\vert $ with $s=\pm 1$. Then, from Eqns. (\ref{inimu}) 
and (\ref{inideriv}), the quantities 
$\mu _{\rm I}(\eta _{\rm i})$ and $\mu _{\rm I}'(\eta _{\rm i})$ take 
the form
\begin{eqnarray}
\label{iniJ1}
\mu _{\rm I}(\eta _{\rm i}) &=& \frac{s^{-1/4}}{\sqrt{2b\gamma }}
\vert \eta _i\vert ^{1/2-b/2}, \\
\label{iniJ2} 
\mu _{\rm I}'(\eta _{\rm i}) &=& \pm is^{1/4}\sqrt{\frac{b\gamma }{2}}
\vert \eta _i\vert ^{-1/2+b/2},
\end{eqnarray}
where we have defined $b$ and $\gamma $ (not to be confused with 
the function $\gamma $ used in Section II) by the following expressions 
\begin{equation}
\label{defbgam}
b\equiv 1-m(1+\beta), \quad \gamma \equiv 
\frac{\sqrt{\vert b_{m}\vert}}{b(2\pi )^{m}}
\epsilon ^{m}n^{m+1}.
\end{equation}
{F}rom the expressions (\ref{iniJ1}) and (\ref{iniJ2}), we deduce 
\begin{equation}
\label{ratio}
\mu '_{\rm I}(\eta _{\rm i})/\mu _{\rm I}(\eta _{\rm i})\, = \, 
\pm is^{1/2}b\gamma \vert \eta _i\vert ^{b-1} \, .
\end{equation}
This ratio will out turn to be important in the calculation 
of the various coefficients determined by the matching 
procedure. To go further, we need to treat the cases $s=\pm 1$ 
separately.

\subsubsection{The case $s=-1$, $b_m<0$}

In Region I, the equation of motion for the mode function reduces to
\begin{equation}
\label{eomJ2}
\mu ''+n^2\frac{b_{m}}{(2\pi )^{2m}}
\biggl(\frac{\epsilon \, n}{\vert \eta \vert ^{1+\beta }}
\biggr)^{2m}\mu =0.
\end{equation}
For a negative coefficient $b_m$, the exact solution of 
Eq. (\ref{eomJ2}) can be expressed in terms 
of modified Bessel functions as follows
\begin{equation}
\label{solJI}
\mu _{\rm I}(\eta )=A_1\vert \eta \vert ^{1/2}
I_{\nu }(z)+A_2\vert \eta \vert ^{1/2}K_{\nu}(z),
\end{equation}
where $\nu \equiv 1/(2b)$ and where the function $z(\eta )$ is defined 
by the following expression 
$z(\eta )\equiv \gamma \vert \eta \vert ^b$. The coefficients $A_1 $ 
and $A_2$ are determined 
by the initial conditions given in Eqns. (\ref{iniJ1}) and (\ref{iniJ2}). These 
coefficients should satisfy the system of equations
\begin{eqnarray}
\label{A1J}
A_1I_{\nu }(z_{\rm i})+A_2K_{\nu }(z_{\rm i})
&=& \vert \eta _{\rm i}\vert ^{-1/2}\mu _{\rm I}(\eta _{\rm i}), \\
\label{A2J}
-A_1I_{\nu -1}(z_i)+A_2K_{\nu -1}(z_i)
&=&\frac{\vert \eta _{\rm i}\vert ^{1/2-b}}{\gamma b}\mu _{\rm I}'(\eta _{\rm i}),
\end{eqnarray}
where $z_{\rm i}$ denotes the value of $z(\eta )$ at time 
$\eta =\eta _{\rm i}$. The exact solution for $A_1$ and $A_2$ can 
be expressed as
\begin{eqnarray}
\label{exactA1}
A_1 &=& \gamma \vert \eta _{\rm i}\vert ^{b-1/2}\mu _{\rm I}(\eta _{\rm i})
K_{\nu -1}(z_{\rm i}) \nonumber \\
& & \times \biggl[1
-\frac{\vert \eta _{\rm i}\vert ^{1-b}}{\gamma b}
\frac{\mu _{\rm I}'(\eta _{\rm i})}{\mu _{\rm I}(\eta _{\rm i})} 
\frac{K_{\nu }(z_{\rm i})}{K_{\nu -1}(z_{\rm i})}\biggr], \\
\label{exactA2}
A_2 &=& \gamma \vert \eta _{\rm i}\vert ^{b-1/2}\mu _{\rm I}(\eta _{\rm i})
I_{\nu -1}(z_{\rm i}) \nonumber \\
& & \times \biggl[1+\frac{\vert \eta _{\rm i}\vert ^{1-b}}{\gamma b}
\frac{\mu _{\rm I}'(\eta _{\rm i})}{\mu _{\rm I}(\eta _{\rm i})} 
\frac{I_{\nu }(z_{\rm i})}{I_{\nu -1}(z_{\rm i})}\biggr].
\end{eqnarray}
In the derivation of the previous expressions, we used the 
exact equation: $(I_{\nu }K_{\nu -1}+I_{\nu -1}K_{\nu })(z)=1/z$. Since, 
when $l_{\rm C}\gg \lambda (\eta _i)$, the argument $z_{\rm i}$ is large we can now 
rewrite these equations using the asymptotic formulas for Bessel functions of 
large arguments \cite{GradR}. Notice that it is necessary to go to the second order 
in the expansion of the modified Bessel functions. We obtain
\begin{eqnarray}
\label{A1approx}
A_1 &\approx & \biggl(\frac{\pi }{2}\biggr)^{1/2}\gamma ^{1/2} \mu _{\rm I}(\eta _{\rm i})
\vert \eta _{\rm i}\vert ^{b/2-1/2}e^{-z_{\rm i}} \nonumber \\
& & \times \biggl[1\pm 1\pm \frac{2\nu -1}{2\gamma }
\vert \eta _{\rm i}\vert ^{-b}\biggr], \\
\label{A2approx}
A_2 &\approx & \biggl(\frac{1}{2\pi }\biggr)^{1/2}\gamma ^{1/2} \mu _{\rm I}(\eta _{\rm i})
\vert \eta _{\rm i}\vert ^{b/2-1/2}e^{z_{\rm i}} \nonumber \\
& & \times \biggl[1\mp 1\mp \frac{1-2\nu }{2\gamma }
\vert \eta _{\rm i}\vert ^{-b}\biggr].
\end{eqnarray}
For the sake of completeness, we pursue the calculation 
for both choices of the sign of the initial 
conditions. Let us again use an index ``u'' for the upper choice and ``l'' for the 
lower choice. We obtain:
\begin{eqnarray}
\label{A1u}
A_1^u &=& 2\biggl(\frac{\pi }{2}\biggr)^{1/2}\gamma ^{1/2} \mu _{\rm I}(\eta _{\rm i})
\vert \eta _{\rm i}\vert ^{b/2-1/2}e^{-z_{\rm i}}, \\
\label{A2u}
A_2^u &=& \biggl(\frac{1}{2\pi }\biggr)^{1/2} \mu _{\rm I}(\eta _{\rm i})
\vert \eta _{\rm i}\vert ^{-b/2-1/2}\frac{2\nu -1}{2\gamma ^{1/2}}e^{z_{\rm i}}, \\
\label{A1l}
A_1^l &=& \biggl(\frac{\pi }{2}\biggr)^{1/2}\mu _{\rm I}(\eta _{\rm i})
\vert \eta _{\rm i}\vert ^{-b/2-1/2}\frac{1-2\nu }{2\gamma ^{1/2}}e^{-z_{\rm i}}, \\
\label{A2l}
A_2^l &=& 2\biggl(\frac{1}{2\pi }\biggr)^{1/2}\gamma ^{1/2} \mu _{\rm I}(\eta _{\rm i})
\vert \eta _{\rm i}\vert ^{b/2-1/2}e^{z_{\rm i}}.
\end{eqnarray}
The exponential factor 
always determines the behaviour of the coefficients for any power 
of $\vert \eta _{\rm i}\vert $. This implies $A_1^u \approx A_1^l\approx 0$. 
We also see the following crucial effect: it turns out that for one choice of 
the sign of the derivative the first term in the squared bracket 
in Eqns. (\ref{A1approx}) and (\ref{A2approx}) cancels whereas for the 
other choice it is no longer the case. This has as a consequence that the dependence 
on $\gamma $ is not the same. Since $\gamma $ depends on $n$, the $n$ dependence 
of $A_2^u$ and $A_2^l$ is not the same. We have $A_2^u\propto \gamma ^{-1}A_2^l$.
\par
The second step of the calculation is to perform the matching 
of the solutions at the time $\eta =\eta _1$. This will allow 
us to express the coefficients $B_1$ 
and $B_2$ in terms of the coefficients $A_1$ and $A_2$. In 
Region II, the solution is given by plane waves. Therefore, the 
coefficients $B_1$ and $B_2$ are now solutions of the equations
\begin{eqnarray}
\label{BJ1}
B_1e^{in\eta _1}+B_2e^{-in\eta _1} &=& 
A_1\vert \eta _1\vert ^{1/2}I_{\nu }(z_1) \nonumber \\
& &+A_2\vert \eta _1\vert ^{1/2}K_{\nu }(z_1), \\
\label{BJ2}
B_1e^{in\eta _1}-B_2e^{-in\eta _1} &=& 
-\frac{\gamma b}{in}A_1\vert \eta _1\vert ^{b-1/2}I_{\nu -1}(z_1) \nonumber \\
& &+ \frac{\gamma b}{in}A_2\vert \eta _1\vert ^{b-1/2}K_{\nu -1}(z_1),
\end{eqnarray}
where $z_1$ is the value of the function $z(\eta )$ at $\eta =\eta _1$. The exact 
solution of this system of equations can be easily found and reads
\begin{eqnarray}
\label{solBJ1}
e^{in\eta _1}B_1 &=& \frac{A_1}{2}\vert \eta _1\vert ^{1/2}
I_{\nu }(z_1)\biggl[1+\frac{i\gamma b }{n}\vert \eta _1\vert ^{b-1}
\frac{I_{\nu -1}(z_1)}{I_{\nu }(z_1)}\biggr] \nonumber \\
& & \hspace*{-1.6cm}+\frac{A_2}{2}\vert \eta _1\vert ^{1/2}
K_{\nu }(z_1)\biggl[1-\frac{i\gamma b}{n}\vert \eta _1\vert ^{b-1}
\frac{K_{\nu -1}(z_1)}{K_{\nu }(z_1)}\biggr], \\
\label{solBJ2}
e^{-in\eta _1}B_2 &=& \frac{A_1}{2}\vert \eta _1\vert ^{1/2}
I_{\nu }(z_1)\biggl[1-\frac{i\gamma b}{n}\vert \eta _1\vert ^{b-1}
\frac{I_{\nu -1}(z_1)}{I_{\nu }(z_1)}\biggr]\nonumber \\
& & \hspace*{-1.6cm}+\frac{A_2}{2}\vert \eta _1\vert ^{1/2}
K_{\nu }(z_1)\biggl[1+\frac{i\gamma b}{n}\vert \eta _1\vert ^{b-1}
\frac{K_{\nu -1}(z_1)}{K_{\nu }(z_1)}\biggr].
\end{eqnarray}
Much simpler (approximate) formulas can be obtained if one notices that the 
argument of the Bessel function is a big number $z_1=\gamma \vert
\eta _1\vert ^b\gg 1$, essentially because $\epsilon $ is a small
number in realistic 
cases. A very simple estimate allows us  to quickly check the 
validity of this approximation. We take $m=1$, 
$\vert b_1\vert =1$, $\beta =-2.2$ which would 
correspond to a spectral index of $n_{\rm S}=0.6$ for power law 
inflation and $\epsilon =10^{-5}$ as already discussed in the 
previous section. We can then 
estimate $z_1$ for $n=4\pi $ which corresponds to the mode 
which re-enters horizon today and which consequently mainly 
determines the value of the CMB quadrupole anisotropy. We find 
$z_1\approx 4.7\times 10^4$. Therefore, we can again use the 
asymptotic behaviour of the Bessel function to simplify the 
previous equations. Putting all these 
ingredients together, we find \footnote{In order to be able 
to neglect the terms proportional to $A_1$, we make use of the 
fact that $|\eta_i| \gg |\eta_1|$.}
\begin{eqnarray}
\label{solBJ1b}
B_1 &\approx & \frac{A_2}{2}\biggl(\frac{\pi }{2\gamma }\biggr)^{1/2} 
\vert \eta _1\vert ^{1/2-b/2}e^{-in\eta _1-z_1-i\frac{\pi }{4}}, \\
\label{solBJ2b}
B_2 &\approx & \frac{A_2}{2}\biggl(\frac{\pi }{2\gamma }\biggr)^{1/2} 
\vert \eta _1\vert ^{1/2-b/2}e^{in\eta _1-z_1+i\frac{\pi }{4}}.
\end{eqnarray}
As a consequence, the solution in Region II can 
be written as
\begin{equation}
\label{solJII}
\mu _{\rm II}(\eta )=A_2\biggl(\frac{\pi }{2\gamma }\biggr)^{1/2}
\vert \eta _1\vert ^{1/2-b/2}e^{-z_1}
\cos\biggl(n\eta -n\eta _1-\frac{\pi }{4} \biggr).
\end{equation}
The last step of the calculation is to perform the matching at 
$\eta =\eta _2$ when the mode leaves the Hubble radius (boundary between Region II 
and Region III). As already 
mentioned, in Region III, the non-decaying solution is the 
super-Hubble function given by 
\begin{equation}
\mu _{\rm III}(\eta )=Ca(\eta ) \, . 
\end{equation}
Repeating the 
same procedure as for Unruh's case, the spectra can easily be
calculated and read 
\begin{eqnarray}
\label{specJl}
n^3P_{\rm \Phi}^l &\propto & n^{2\beta+4}e^{2(z_{\rm i}-z_1)}
\cos ^2\biggl(n\eta _2-n\eta _1-\frac{\pi }{4} \biggr), \\ 
\label{specJu}
n^3P_{\rm \Phi}^u &\propto & n^{2\beta+2-2m}e^{2(z_{\rm i}-z_1)}
\cos ^2\biggl(n\eta _2-n\eta _1-\frac{\pi }{4} \biggr).
\end{eqnarray}
We see that the spectrum depends explicitly on the initial 
conditions chosen. We can check that the tilt is correct by noticing 
that $n^3P_{\rm \Phi }\propto A_2^2$, $\gamma \propto n^{m+1}$ and using the 
relation between $A_2^u$ and $A_2^l$ already mentioned. From now on, we 
concentrate on the lower case which corresponds to an unmodified 
tilt and study 
the expression of the corresponding spectrum in more 
details (for convenience, we drop the subscript `l'). First, as 
mentioned above, we see that the power-law part is not modified in 
comparison with the usual case, i.e. the spectral index is still 
$n_{\rm S}=2\beta +5$. Secondly, there are 
oscillations in the spectrum since the 
argument of the cosine can be written as (considering for simplicity
that $\vert b_m\vert =1$)
\begin{equation}
\label{arg}
2\pi |1+\beta |-\biggl(\frac{\epsilon }{2\pi }\biggr)^{1/(1+\beta )}
n^{(2+\beta )/(1+\beta )}-\frac{\pi }{4} .
\end{equation}
However, contrary to Unruh's case with Minkowski initial conditions, no logarithmic 
dependence is present. Interestingly enough, for $\beta =-2$, the oscillations 
disappear. The most important part concerns the exponential factor. The factor 
$z_{\rm i}-z_1$ is equal to $z_{\rm i}-z_1 =
\gamma |\eta _{\rm i}|^b(1-|\eta _1|^b/|\eta _{\rm i}^b)
\approx \gamma |\eta _{\rm i}|^b=z_{\rm i}$ since we 
have $|\eta _{\rm i}|\gg |\eta _1|$. The factor $z_{\rm i}$ can be re-written 
in such a way that the dependence on $n$ is explicit
\begin{equation}
\label{expfac}
z_{\rm i}=\frac{\sqrt{\vert b_{m}\vert}}{b(2\pi )^m}
\epsilon ^m|\eta _{\rm i}|^{1-m(1+\beta )}n^{m+1}.
\end{equation}
The important factor in this expression is 
$\epsilon ^m|\eta _{\rm i}|^{1-m(1+\beta )}$ since the others ones are 
of order one. It can be re-written as 
\begin{equation}
\label{impofac}
\epsilon ^m|\eta _{\rm i}|^{1-m(1+\beta )}
=\biggl[\frac{l_{\rm C}}{a(\eta _{\rm i})}\biggr]^m|\eta _{\rm i}|,
\end{equation}
and must be considered as large since $|\eta _{\rm i}|\gg 1 $ and 
$l_{\rm C}/a(\eta _{\rm i})\approx l_{\rm C}/\lambda (\eta _{\rm i})\gg 1$, 
at least for wavenumbers not too different from $2\pi $. This means 
that the influence of the exponential factor is dominant and is responsible 
for a huge increase of the spectrum at large $n$. This is illustrated if 
we write the spectrum for $\beta =-2$ and $m=1$
\begin{equation}
n^3P_{\rm \Phi}\propto e^{An^2},
\end{equation}
where $A\gg 1$. Such a spectrum is almost certainly in contradiction 
with observations.
\par
We end this subsection with the calculation of the spectrum in the case 
where the initial state is the minimum energy density state. We restart 
from the exact expressions for the coefficients $A_1$ and $A_2$, see 
Eqns. (\ref{exactA1}) and (\ref{exactA2}). Using Eqns. (\ref{Minvac}), 
we have 
\begin{equation}
\label{JCMin}
\frac{\vert \eta _{\rm i}\vert ^{1-b}}{\gamma b}\frac{\mu _{\rm I}'(\eta _{\rm i})}
{\mu _{\rm I}(\eta _{\rm i})}=\pm \frac{i}{\sqrt{b_{m}}}
\biggl[\frac{\lambda (\eta _{\rm i})}{l_{\rm C}}\biggr]^m \ll 1, 
\end{equation}
since, initially, $l_{\rm C}\gg \lambda (\eta _{\rm i})$. As a consequence, 
we can derive a compact approximate expression for the coefficients $A_1$ 
and $A_2$
\begin{equation}
\label{AminkJC}
A_1\approx 0, \quad A_2\approx \frac{1}{\sqrt{2\pi }}\gamma ^{1/2}\vert \eta _{\rm i}
\vert ^{b/2-1/2}\mu _{I}(\eta _{\rm i})e^{z_{\rm i}}.
\end{equation}
Notice that these formulas are valid for any choice of the 
sign of $\mu _{\rm I}'(\eta _{\rm i})$. The rest of the calculation 
proceeds as above and leads to ($\vert b_m\vert =1$)
\begin{eqnarray}
\label{PminkJC}
n^3P_{\rm \Phi} &=& n^{2\beta +4+m}e^{An^{m+1}} \nonumber \\
& &\times \cos ^2
\biggl[2\pi |1+\beta |-\biggl(\frac{\epsilon }{2\pi }\biggr)^{\frac{1}{1+\beta }}
n^{\frac{2+\beta }{1+\beta }}-\frac{\pi }{4} \biggr].
\end{eqnarray}
The main difference in comparison with the spectrum of the previous section is the 
presence of a modified tilt. The spectral index is now given 
by $n_{\rm S}=2\beta +5+m$.

\subsubsection{The case $s=1$, $b_m>0$}

When the dispersion relation is real, the solution in Region I can be 
expressed in terms of usual Bessel functions
\begin{equation}
\label{sol}
\mu _{\rm I}(\eta )=A_1\vert \eta \vert ^{1/2}J_{\nu }(z)+
A_2\vert \eta \vert ^{1/2}J_{-\nu }(z),
\end{equation}
where $\nu $ and $z(\eta )$ have already been defined previously. The 
coefficients $A_1$ and $A_2$ are now solutions of the following system 
of equations
\begin{eqnarray}
\label{Ai1}
A_1J_{\nu }(z_{\rm i})+A_2J_{-\nu }(z_{\rm i}) &=& \mu _{\rm I}(\eta _{\rm i})
\vert \eta _{\rm i}\vert ^{-1/2}, \\
\label{Ai2}
-A_1J_{\nu }(z_{\rm i})+A_2J_{-\nu }(z_{\rm i}) &=& 
\frac{\mu _{\rm I}'(\eta _{\rm i})}{\gamma b}\vert \eta _{\rm i}\vert ^{1/2-b}.
\end{eqnarray}
Using the relation expressing the Wronskian $[J_{-\nu }J_{\nu -1}
+J_{-\nu +1 }J_{\nu }](z)= 
2\sin [\pi/(2b)]/(\pi z)$, and performing some straightforward 
algebraic manipulations, exact expressions can be easily found. They read 
\begin{eqnarray}
\label{Ai1sol}
A_1 &=& \frac{\pi \gamma }{2\sin (\pi \nu )} \vert \eta _{\rm i}\vert ^{b-1/2}
\mu _{\rm I}(\eta _{\rm i})J_{1-\nu }(z_{\rm i}) \nonumber \\
& & \times \biggr[1-
\frac{\vert \eta _{\rm i}\vert ^{1-b}}{\gamma b}
\frac{\mu _{\rm I}'(\eta _{\rm i})}{\mu _{\rm I}(\eta _{\rm i})}
\frac{J_{-\nu }(z_{\rm i})}{J_{1-\nu }(z_{\rm i})}\biggr], \\
\label{Ai2sol}
A_2 &=& \frac{\pi \gamma }{2\sin (\pi \nu )} \vert \eta _{\rm i}\vert ^{b-1/2}
\mu _{\rm I}(\eta _{\rm i})J_{\nu -1}(z_{\rm i}) \nonumber \\
& & \times 
\biggr[1+
\frac{\vert \eta _{\rm i}\vert ^{1-b}}{\gamma b}
\frac{\mu _{\rm I}'(\eta _{\rm i})}{\mu _{\rm I}(\eta _{\rm i})}
\frac{J_{\nu }(z_{\rm i})}{J_{\nu -1}(z_{\rm i})}\biggr].
\end{eqnarray}
These expressions are not valid if $\nu =1/(2b)$ is an integer and 
this particular case must be treated separately. In this article, we 
assume that this does not happen. Since the solution in Region II is still given 
by plane waves, the derivation of exact expressions for the coefficients $B_1$ 
and $B_2$ proceeds as before. Explicit matching of the mode 
function and of its derivative leads to
\begin{eqnarray}
\label{B1}
e^{in\eta _1}B_1 &=& \frac{A_1}{2}\vert \eta _1\vert ^{1/2}J_{\nu }(z_1)
\biggl[1+i\frac{\gamma b}{n}\vert \eta _1\vert ^{b-1}
\frac{J_{\nu -1}(z_1)}{J_{\nu }(z_1)}\biggr] \nonumber \\
& & \hspace*{-1.5cm}+\frac{A_2}{2}\vert \eta _1\vert ^{1/2}J_{-\nu }(z_1)
\biggl[1-i\frac{\gamma b}{n}\vert \eta _1\vert ^{b-1}
\frac{J_{-\nu +1}(z_1)}{J_{-\nu }(z_1)}\biggr], \\
\label{B2}
e^{-in\eta _1}B_2 &=& \frac{A_1}{2}\vert \eta _1\vert ^{1/2}J_{\nu }(z_1)
\biggl[1-i\frac{\gamma b}{n}\vert \eta _1\vert ^{b-1}
\frac{J_{\nu -1}(z_1)}{J_{\nu }(z_1)}\biggr] \nonumber \\
& & \hspace*{-1.5cm}+\frac{A_2}{2}\vert \eta _1\vert ^{1/2}J_{-\nu }(z_1)
\biggl[1+i\frac{\gamma b}{n}\vert \eta _1\vert ^{b-1}
\frac{J_{-\nu +1}(z_1)}{J_{-\nu }(z_1)}\biggr].
\end{eqnarray}
Having all the relevant exact expressions at our disposal we 
can now start to do some approximations based on the fact that 
$z_{\rm i}$ is a big number. For convenience, we introduce two 
new definitions [not to be confused with the functions $x(\eta )$ and 
$y(\eta )$ introduced in section IV-B]
\begin{equation}
\label{defxy}
x(\eta )\equiv z(\eta )+\frac{\pi \nu }{2}-\frac{\pi }{4}, \quad 
y(\eta )\equiv z(\eta )-\frac{\pi \nu }{2}-\frac{\pi }{4}.
\end{equation}
Then, using the expressions of the Bessel functions for 
large arguments \cite{GradR}, we find 
\begin{eqnarray}
\label{A1approxJCr}
A_1 &\approx & \mp i\biggl(\frac{\pi \gamma }{2}\biggr)^{1/2}
\vert \eta _{\rm i}\vert ^{b/2-1/2}
\frac{\mu _{\rm I}(\eta _{\rm i})}{\sin (\pi \nu )}e^{\pm ix_{\rm i}}, \\
\label{A2approxJCr}
A_2 &\approx & \pm i\biggl(\frac{\pi \gamma }{2}\biggr)^{1/2}
\vert \eta _{\rm i}\vert ^{b/2-1/2}
\frac{\mu _{\rm I}(\eta _{\rm i})}{\sin (\pi \nu )}e^{\pm iy_{\rm i}},
\end{eqnarray}
where $x_{\rm i}\equiv x(\eta _{\rm i})$ and 
$y_{\rm i}\equiv y(\eta _{\rm i})$. The correct matching time is 
$\vert \eta _1\vert =[n\epsilon /(2\pi )]^{1/(1+\beta
)}b_m^{1/[2m(1+\beta )]}$, see Ref. \cite{BM4}, and is equal to the
time at which $\lambda =l_{\rm C}$ if $b_m=1$. In the following, for 
simplicity, we consider $b_m=1$. The coefficients 
$B_1$ and $B_2$ can be expressed as
\begin{eqnarray}
\label{B1approxJCr}
B_1 &\approx & \biggl(\frac{1}{2\pi \gamma }\biggr)^{1/2}\vert \eta _1\vert ^{1/2-b/2}
e^{-in\eta _1}\biggl(A_1\cos y_1 \nonumber \\
& &\hspace*{-0.7cm} -iA_1\sin y_1
+A_2\cos x_1-iA_2\sin x_1\biggr), \\
\label{B2approxJCr}
B_2 &\approx & \biggl(\frac{1}{2\pi \gamma }\biggr)^{1/2}\vert \eta _1\vert ^{1/2-b/2}
e^{in\eta _1}\biggl(A_1\cos y_1 \nonumber \\
& &\hspace*{-0.7cm} +iA_1\sin y_1
+A_2\cos x_1+iA_2\sin x_1\biggr),
\end{eqnarray}
where $x_1\equiv x(\eta _1)$ and $y_1=y(\eta _1)$. Our next move is to 
replace the expressions of $A_1$ and $A_2$, see Eqns. (\ref{A1approxJCr}) and 
(\ref{A2approxJCr}), in the previous formula. This leads to
\begin{eqnarray}
\label{B1fin}
B_1 &=& \mp i\frac{\mu _{\rm I}(\eta _{\rm i})e^{-in\eta _1}}{2\sin (\pi \nu )}
\biggl\vert \frac{\eta _1}{\eta _{\rm i}}\biggr \vert ^{1/2-b/2}e^{\pm ix_{\rm i}}
\biggl(\cos y_1 \nonumber \\
& & \hspace*{-0.9cm}-i\sin y_1-e^{\mp i\pi \nu}\cos x_1
+ie^{\mp i\pi \nu}\sin x_1\biggl), \\
\label{B2fin}
B_2 &=& \mp i\frac{\mu _{\rm I}(\eta _{\rm i})e^{in\eta _1}}{2\sin (\pi \nu )}
\biggl\vert \frac{\eta _1}{\eta _{\rm i}}\biggr \vert ^{1/2-b/2}e^{\pm ix_{\rm i}}
\biggl(\cos y_1 \nonumber \\
& & \hspace*{-0.9cm}+i\sin y_1-e^{\mp i\pi \nu}\cos x_1
-ie^{\mp i\pi \nu}\sin x_1\biggl).
\end{eqnarray}
Then, the mode function at time $\eta =\eta _2$ (which is the relevant 
quantity for the determination of the constant $C$) can be expressed as
\begin{equation}
\label{mu2}
\mu _{\rm II}(\eta _2)=\mp i\frac{\mu _{\rm I}(\eta _{\rm i})}{2\sin (\pi \nu )}
\biggl\vert \frac{\eta _1}{\eta _{\rm i}}\biggr \vert ^{1/2-b/2}e^{\pm ix_{\rm i}}.
\end{equation}
{F}rom this equation, the expression of the spectrum 
can be easily established and reads
\begin{equation}
\label{specJCr}
n^3P_{\rm \Phi }\propto n^{2\beta +4}.
\end{equation}
Let us analyze this spectrum in more detail. The first 
remark is that the tilt is unchanged and that the spectral 
index is given by the usual expression $n_{\rm S}=2\beta +5$. 
The second remark is that the exponential dependence has 
disappeared. This is due to the fact that, for $s=1$, this 
factor becomes a pure phase. We recover the usual result 
as pointed out in Ref. \cite{NPA}. 
\par
Let us finally turn to the case where the initial conditions 
are those which correspond to the instantaneous Minkowski vacuum. 
Restarting from the exact expressions for the coefficients 
$A_1$ and $A_2$, see Eqns. (\ref{Ai1sol}) and (\ref{Ai2sol}), and using 
Eqns. (\ref{Minvac}), we find
\begin{eqnarray}
\label{A1approxmin}
A_1 &\approx & \biggl(\frac{\pi \gamma }{2}\biggr)^{1/2}
\vert \eta _{\rm i}\vert ^{b/2-1/2}
\frac{\mu _{\rm I}(\eta _{\rm i})}{\sin (\pi \nu )}\sin x_{\rm i}, \\
\label{A2approxmin}
A_2 &\approx & -\biggl(\frac{\pi \gamma }{2}\biggr)^{1/2}
\vert \eta _{\rm i}\vert ^{b/2-1/2}
\frac{\mu _{\rm I}(\eta _{\rm i})}{\sin (\pi \nu )}\sin y_{\rm i}.
\end{eqnarray}
Inserting these equations into the exact formulas giving the coefficients 
$B_1$ and $B_2$, we obtain 
\begin{eqnarray}
\label{B1approxmin}
B_1 &\approx & \frac{\mu _{\rm I}(\eta _{\rm i})e^{-in\eta _1}}{2\sin (\pi \nu )}
\biggl \vert \frac{\eta _1}{\eta _{\rm i}}\biggr \vert ^{-b/2+1/2} \nonumber \\
& & \times \biggl(\sin x_{\rm i}\cos y_1 
-i\sin x_{\rm i}\sin y_1 -\sin y_{\rm i}\cos x_1 \nonumber \\
& & +i\sin y_{\rm i}\sin x_1\biggr), \\
\label{B2approxmin}
B_2 &\approx & \frac{\mu _{\rm I}(\eta _{\rm i})e^{in\eta _1}}{2\sin (\pi \nu )}
\biggl \vert \frac{\eta _1}{\eta _{\rm i}}\biggr \vert ^{-b/2+1/2} \nonumber \\
& & \times \biggl(\sin x_{\rm i}\cos y_1 
+i\sin x_{\rm i}\sin y_1 
-\sin y_{\rm i}\cos x_1 \nonumber \\
& &-i\sin y_{\rm i}\sin x_1\biggr).
\end{eqnarray}
We are now in a position where we can write the expression of the 
mode function at time $\eta =\eta _2$. It reads
\begin{eqnarray}
\label{muII}
\mu _{\rm II}(\eta _2) &=& \frac{\mu _{\rm I}(\eta _{\rm i})}{2\sin (\pi \nu )}
\biggl \vert \frac{\eta _1}{\eta _{\rm i}}\biggr \vert ^{-b/2+1/2}
\biggl[\tilde{B}(n)e^{in(\eta _2-\eta _1)} \nonumber \\
& & +\tilde{B}^*(n)e^{-in(\eta _2-\eta _1)}\biggr],
\end{eqnarray}
where the function $\tilde{B}(n)$ is defined by
\begin{eqnarray}
\label{tildeB}
\tilde{B}(n) &\equiv &\sin x_{\rm i}\cos y_1-i\sin x_{\rm i}\sin y_1
-\sin y_{\rm i}\cos x_1 \nonumber \\
& & +i\sin y_{\rm i}\sin x_1.
\end{eqnarray}
Then, one can write $\tilde{B}(n)$ as $\tilde{B}(n)\equiv \vert 
\tilde{B} \vert e^{i\psi}$ and define $\bar{B}(n)$ as 
$\bar{B}(n)\equiv \vert \tilde{B}(n)\vert \cos (n\eta _2-n\eta _1+\psi)$. It 
follows that the spectrum can be written as
\begin{equation}
\label{specmink}
n^3P_{\rm \Phi }\propto n^{2\beta +4+m}\vert \bar{B}(n)\vert ^2.
\end{equation}
The spectral index is given by $n_{\rm S}=2\beta +5+m$, i.e. it differs 
from the standard one but is equal to the spectral index obtained 
in the case $s=-1$ for instantaneous Minkowski initial 
conditions. The factor $\vert \bar{B}(n)\vert ^2$ is of order one and 
produces a complicated oscillatory pattern.
\par
In conclusion, the resulting spectrum in the case of the Corley/Jacobson 
dispersion relation is very different from the usual spectrum calculated 
using an unmodified dispersion relation, and different from what is obtained 
using Unruh's relation, even for initial conditions which minimize the energy. 

\section{Discussion and Conclusions}

We have studied the dependence of the predictions of inflationary cosmology for 
the spectrum of fluctuations on hidden assumptions about super-Planck-scale 
physics. The motivation for our work is that in most current models of 
inflation, the period of exponential expansion lasts so long that at the beginning 
of inflation, scales of cosmological interest today had a physical wavelength much 
smaller than the Planck length, and the theories used to compute the spectrum of 
fluctuations are known to break down on these scales.
\par
We studied the problem by replacing the dispersion relation of a free field 
theory which is used to compute the spectrum in the standard approaches by a 
modified dispersion relation, the modifications only being important on length 
scales smaller than a cutoff length $l_{\rm C}$ (which we expect to be given by the 
Planck length). We considered two classes of dispersion relations, based on the 
ones considered by Unruh \cite{Unruh} (Class A) and by Corley and 
Jacobson \cite{CJ} (Class B), respectively, in their studies of the trans-Planckian 
problem of black hole physics. Admittedly, modifying the physics in this way is a 
very ad hoc way of taking into account possible effects of super-Planck-scale 
physics, chosen for mathematical simplicity. We do not want to introduce mode-mode 
coupling in order to keep the computations simple. However, in order to demonstrate 
that there is a possible problem for the robustness of the usual predictions of 
inflation it is sufficient to construct one example of a modified theory which 
leads to different predictions.
\par
For a non-standard dispersion relation the choice of initial state becomes more 
difficult. We considered two choices, both of which coincide with the usual 
initial state in the case of the standard dispersion relation. The first (and 
better motivated) choice is the state which minimizes the energy density, the 
second choice is the naive generalization of the `local Minkowski vacuum'.
\par
We have shown that for Class A dispersion relations the usual predictions of 
inflationary cosmology are recovered (in the case of exponential inflation)
if the initial state minimizes the energy density. In particular, the
spectrum of fluctuations is scale-invariant. If the initial state is chosen
to be the `local Minkowski vacuum', then the resulting spectrum has a tilt 
and superimposed oscillations. 
\par
In contrast, for Class B dispersion relations and an initial state which 
minimizes the energy density, the resulting spectrum of fluctuations is in
general {\bf not} scale-invariant. The precise nature of the spectrum depends 
sensitively on whether the dispersion relation turns complex or remains real. In 
the complex case, the spectrum is characterized by an exponential factor (more 
power in the blue, i.e. $n_{\rm S} > 1$), a tilt (compared to the ``standard" 
predictions) which depends on the precise initial conditions, and superimposed 
oscillations. The exponent, the tilt, and the precise oscillatory pattern depend 
on the specific member of the class of dispersion relations chosen. For a 
spectrum which remains real, the usual result is unchanged.  
\par
The reason why for Class A dispersion relations the usual predictions of 
inflation are maintained is that the time evolution during the period when 
the mode wavelength is smaller than the cutoff scale is adiabatic. This emerges
from our calculations, but an intuitive way of understanding the 
result is that at all times the effective frequency of the mode is larger 
than the Hubble rate and the initial vacuum state therefore adjusts itself 
adiabatically to track the instantaneous vacuum state, thus
leading to the same state at time $\eta_1$ as in the theory with unmodified
dispersion relation \footnote{We thank Bill Unruh for making this point 
to us.}. For Class B dispersion relations, in contrast, the dispersion
relation varies too quickly as a function of time while the scale is smaller 
than the critical length $l_{\rm C}$ and hence the evolution is not adiabatic 
.
\par
Let us now be more quantitative about the previous discussion. In Region I, 
Eq. (\ref{eqommu}) can be written as
\begin{equation}
\label{eqmotI}
\mu ''+n_{\rm eff}^2\mu =\mu ''+a^2(\eta )\omega ^2_{\rm phys}(n,\eta )\mu =0,
\end{equation}
where $\omega _{\rm phys}$ is the physical frequency defined by 
$\omega _{\rm phys}\equiv (1/a)\sqrt{n^2+
a^2\Omega ^2(n,\eta )/l_{\rm C}^2}$. The latter can be considered as constant as 
long as its characteristic time scale of evolution is small compared to 
the Hubble time, i.e. as long as we have adiabaticity. Therefore, let us 
define an ``adiabaticity coefficient'' $\alpha $ according to
\begin{equation}
\label{defalpha}
\alpha (n,\eta )\equiv \Biggl\vert \frac{{\cal H}}{\frac{1}{\omega _{\rm phys }}
\frac{{\rm d}\omega _{\rm phys }}{{\rm d}\eta }}\Biggr\vert,
\end{equation}
where we recall that ${\cal H}\equiv a'/a$. When $\alpha \gg 1$, adiabaticity 
is satisfied and Eq. (\ref{eqmotI}) reduces 
to the equation of motion in the Unruh's case, Eq. (\ref{eomUI}). In this 
situation, we know that the final spectrum is unmodified since there is 
an exact cancellation of the $n$-dependence in the minimizing energy 
state and in the growth factor before Hubble radius crossing which 
results in the usual spectrum. The previous 
argument shows that an unmodified spectrum is expected when $\alpha \gg 1 $ 
in the region where the dispersion relation is modified. Let us also note in 
passing that for the standard 
case, $\alpha =1$, since the time scale of evolution of 
$\omega _{\rm phys}(n,\eta )$ and of the Hubble rate is the same. 
\par
\begin{figure}
\begin{center}
\leavevmode
\hspace*{-2.cm}
\epsfxsize=11cm
\epsffile{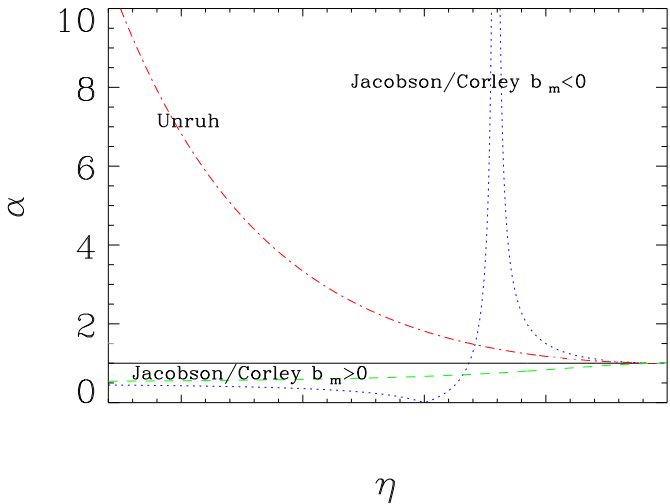}
\end{center}
\caption{Sketch of the time evolution of the adiabaticity coefficient $\alpha $ 
for the dispersion relations.}
\label{adia}
\end{figure}

We have calculated the adiabaticity coefficient for the different 
cases treated in this article. The result is displayed in 
Fig. (\ref{adia}). When $\eta $ goes to $-\infty $, we have 
$l_{\rm C}\gg \lambda $ whereas $l_{\rm C}\ll \lambda $ when 
$\eta $ goes to zero.
\par
We see that there exists a clear difference between Unruh's case and 
the Corley/Jacobson cases. In the Unruh's case, $\alpha $ goes to 
infinity when $l_{\rm C}\gg \lambda $ and adiabaticity is preserved. 
When $b_m<0$, the 
adiabaticity coefficient reaches zero at the time when $\omega =0$ 
. Then, adiabaticity 
is progressively re-established. The coefficient $\alpha $ 
goes to infinity and there is a divergence when ${\rm d}\omega 
/{\rm d}\eta=0$ 
. In the regime when 
$l_{\rm C}\ll \lambda $, $\alpha $ goes to one as it should since the 
various dispersion relations all become similar to the standard 
one. The previous considerations explain why the final spectrum can be 
modified in the Corley/Jacobson case with a complex dispersion 
relation but not in Unruh's case.

We conclude that it is possible that in models of inflation based consistently 
on a unified theory at the Planck scale the predictions for fluctuations will 
not coincide with the usual predictions from our current inflationary Universe 
models. Generalizing from our results, a crucial issue appears to be whether 
the evolution of the quantum states corresponding to the fluctuations will be 
adiabatic on length scales smaller than the Planck scale.
\par
Our results point to the possibility that the interaction between fundamental 
physics and cosmology may be much richer than hitherto assumed. It is not 
only a question of {\it if} and {\it how} fundamental physics leads to 
inflation; a much richer question is {\it what} the specific predictions 
of the {\it fundamental model} of inflation will be, assuming for the sake 
of argument that such a {\it fundamental model} exists. 
\par
It is not surprising that super-Planck-scale physics may modify the usual 
predictions of inflation. One model of the early Universe motivated by string 
theory, the {\it Pre-Big-Bang Cosmology} \cite{PBB} based on dilaton gravity, leads 
to a super-exponential period of early evolution in which the Hubble constant 
is increasing, and where the predicted spectrum of scalar metric fluctuations 
is not scale-invariant \cite{PBBflucts}. It would be interesting to analyze 
the predictions of other models of inflation based on string theory, taking 
into account the evolution on string scales. One toy model in which this question 
could be analyzed is the {\it nonsingular Universe} \cite{MB92} based on 
higher derivative terms in the gravitational action.
\par
In the context of the models studied here, it would be interesting to explore 
whether the {\it minimum energy density} initial state is an attractor in a 
similar sense that the {\it local Minkowski vacuum} is in 
standard inflationary cosmology \cite{BH85}.
\par
Note that models of inflation based on a strongly interacting theory (such as 
the model analyzed in \cite{BZ97}) do not suffer from the Trans-Planckian 
problem discussed in this paper. In strongly interacting theories, perturbations 
are generated at all times at a fixed physical scale, and a scale-invariant 
spectrum results based on the heuristic arguments mentioned in the 
Introduction. In such theories, however, the presence of strong interactions 
makes it hard to calculate the amplitude of the resulting spectrum.

\centerline{\bf Acknowledgements}

We are grateful to Lev Kofman, Dominik Schwarz, Carsten Van de Bruck 
and in particular Bill Unruh for 
stimulating discussions and useful comments. We also thank an
anonymous referee for useful comments. We acknowledge support 
from the BROWN-CNRS University Accord which made possible the visit of J.~M. to 
Brown during which most of the work on this project was done, and we are 
grateful to Herb Fried for his efforts to secure this Accord. One of us (R.~B.) 
wishes to thank Bill Unruh for hospitality at the University of British Columbia 
during the time when this work was completed. J.~M. thanks the High Energy 
Group of Brown University for warm hospitality. The research was supported in 
part by the U.S. Department of Energy under Contract DE-FG02-91ER40688, TASK A.

\end{document}